\documentclass[aps,prl,reprint,twocolumn,showpacs,preprintnumbers,amsmath,amssymb]{revtex4-2}
\usepackage[colorlinks=true,citecolor=blue,urlcolor=blue,linkcolor=blue,bookmarksopen]{hyperref}
\usepackage{amsmath}
\usepackage{amssymb}
\usepackage{mathrsfs}
\usepackage{graphicx}
\usepackage{bm}
\usepackage{comment}
\usepackage{latexsym}
\usepackage{dcolumn}
\usepackage{amsmath}
\usepackage{epsf}
\usepackage{float}
\usepackage[dvipsnames]{xcolor}

\usepackage{natbib}
\usepackage{bm}
\usepackage{mathrsfs}

\usepackage[export]{adjustbox}
\usepackage{amsfonts}
\usepackage{amsmath}
\usepackage[makeroom]{cancel}
\usepackage[english]{babel}
\usepackage[T1]{fontenc}
\usepackage{times}
\usepackage{mathrsfs}
\usepackage{graphicx}
\usepackage{dcolumn}
\usepackage{bm}
\usepackage{wasysym}
\usepackage[tight, FIGTOPCAP, hang, raggedright, nooneline]{subfigure}
\usepackage{xcolor}
\usepackage{epsfig}
\usepackage{lipsum}
\usepackage{enumitem}

\def\d{\partial}
\def\dd{\text{d}}

\def\>{\rangle}
\def\<{\langle}

\renewcommand\d{\partial}

\newcommand\+{\dagger}

\newcommand{\be}{\begin{equation}}
\newcommand{\ee}{\end{equation}}
\newcommand{\bea}{\begin{eqnarray}}
\newcommand{\eea}{\end{eqnarray}}

\newcommand{\bigzero}{\mbox{\normalfont\Large\bfseries 0}}

\newcommand{\II}{\mathbb{I}}

\newcommand{\abs}[1]{\left| #1\right|}
\renewenvironment{widetext@grid}{%
	\par\ignorespaces
	\setbox\widetext@top\vbox{%
		\vskip15\p@
		\hb@xt@\hsize{%
			\leaders\hrule\hfil
			\vrule\@height6\p@
		}%
		\vskip6\p@
	}%
	\setbox\widetext@bot\hb@xt@\hsize{%
		\vrule\@depth6\p@
		\leaders\hrule\hfil
	}%
	\onecolumngrid
	\let\set@footnotewidth\set@footnotewidth@ii
}{%
	\par
	\twocolumngrid\global\@ignoretrue
	\@endpetrue
}%
\makeatother

\begin{document}
	\title{Magic configurations in Moir\'e Superlattice of Bilayer Photonic crystal: \\  Almost-Perfect Flatbands and Unconventional Localization  }
	\author{Dung Xuan Nguyen$^1$}
	\email{dung\_x\_nguyen@brown.edu}	 
	\author{Xavier Letartre$^2$}
	\author{Emmanuel Drouard$^2$}
	\author{Pierre Viktorovitch$^2$}
	\author{H. Chau Nguyen$^3$}
	\author{Hai Son Nguyen$^{2,4}$}
	\email{hai_son.nguyen@ec-lyon.fr}
	\affiliation{$^1$Brown Theoretical Physics Center and Department of Physics, Brown University, 182 Hope Street, Providence, Rhode Island 02912, USA}   
	\affiliation{$^2$Univ Lyon, Ecole Centrale de Lyon, CNRS, INSA Lyon, Universit\'e Claude Bernard Lyon 1, CPE Lyon, CNRS, INL, UMR5270, 69130 Ecully, France} 
	\affiliation{$^3$Naturwissenschaftlich-Technische Fakult\"at, Universit\"at Siegen, Walter-Flex-Stra{\ss}e 3, 57068 Siegen, Germany} 
    \affiliation{$^4$Institut Universitaire de France (IUF)}
	\date{\today}		
	\begin{abstract}
		We investigate the physics of photonic band structures of the moir\'e patterns that emerged when overlapping two uni-dimensional (1D) photonic crystal slabs with mismatched periods. The band structure of our system is a result of the interplay between intra-layer and inter-layer coupling mechanisms, which can be fine-tuned via the distance separating the two layers. 
		We derive an effective Hamiltonian that captures the essential physics of the system and reproduces all numerical simulations of electromagnetic solutions with high accuracy. 
		Most interestingly, \textit{magic distances} corresponding to the emergence of photonic flatbands within the whole Brillouin zone of the moir\'e superlattice are observed. 
		We demonstrate that these flatband modes are tightly localized within a moir\'e period. Moreover, we suggest a single-band tight-binding model that describes the moir\'e minibands, of which the tunnelling rate can be continuously tuned via the inter-layer strength. 
		Our results show that the band structure of bilayer photonic moir\'e can be engineered in the same fashion as the electronic/excitonic counterparts. It would pave the way to study many-body physics at photonic moir\'e flatbands and novel optoelectronic devices.
	\end{abstract}
	\maketitle
	\makeatletter
	\let\toc@pre\relax
	\let\toc@post\relax
	\makeatother         
	
	
	Moir\'{e} structures have been of central interest in fundamental physics during the last few years. The most important milestone is the discovery of flatbands in the moir\'e patterns emerged when two graphene layers are overlapped at certain at \textit{magic} twisted angles\cite{Bistritzer:2010,Tarnopolsky2019,Lisi2021}, leading to non-conventional superconductivity\cite{Cao2018,Arora2020,Stepanov2020} and strongly correlating insulator states with nontrivial-topology\cite{Song2019,wu2020chern}. Motivated by the electronic \textit{magic angles}, photonic moir\'e has attracted tremendous research in light of shaping novel optical phenomena.  Hu \textit{et al.} have demonstrated \cite{Hu2020a,Hu2020b} the topological transition of photonic dispersion in  twisted 2D materials. However, the operating wavelength in these pioneering works are  much larger than the moir\'e period, thus dispersion engineering is based on the anisotropy of an effective medium rather than the microscopic moir\'e pattern. On the other hand,  Ye's group has recently reported on the realization of 2D photonic moir\'e superlattice\cite{Wang2020}. Nevertheless, this work only focused on light scattering through the moir\'e pattern, but the lattice is on the same plane, and there is no bi-layer, neither twisting concepts. Most recently, numerical\cite{Lou2021} and tight-binding\cite{Dong2021} method have been proposed to investigate twisted bilayer photonic crystal slabs. In particular, Dong \textit{et al.} has showed that local flatband would be achieved\cite{Dong2021} in twisted bilayer photonic crystal at small twisted angle..

	In this work, we report on a theoretical study of photonic band structures in moir\'e patterns that emerged when two mismatched 1D subwavelength photonic crystal slabs are overlapped. The essential physics of the system can be captured by an effective four-component Hamiltonian. Accompanying the analytical theory, numerical electromagnetic simulations are performed with a case study of silicon structures operating at telecom wavelength. The obtained band structure are resulted from an interplay between intra-layer and inter-layer coupling mechanisms which is tuned via the distance separating the two layers. Importantly, \textit{magic distances} corresponding to the emergence of photonic flatbands within the whole Brillouin zone are demonstrated. The minibands of moir\'e superlattice can be described by a single-band tight-binding model with Wannier functions tightly confined within a moir\'e period. The tunnelling rate of light between nearest neighbor Wannier states is continuously modulated by the inter-layer distance and vanished at \textit{magic distance}, leading to flatband formation and photonic localization. Despite its simplicity, this 1D setup captures much interesting physics of moir\'{e} systems of twisted two-dimensional materials. Our findings suggest that moir\'e photonic is a promising strategy to engineer photonic bandstructure for fundamental research and optoelectronic devices. 
	
	\begin{figure}
		\begin{center}
			\includegraphics[width=0.45 \textwidth]{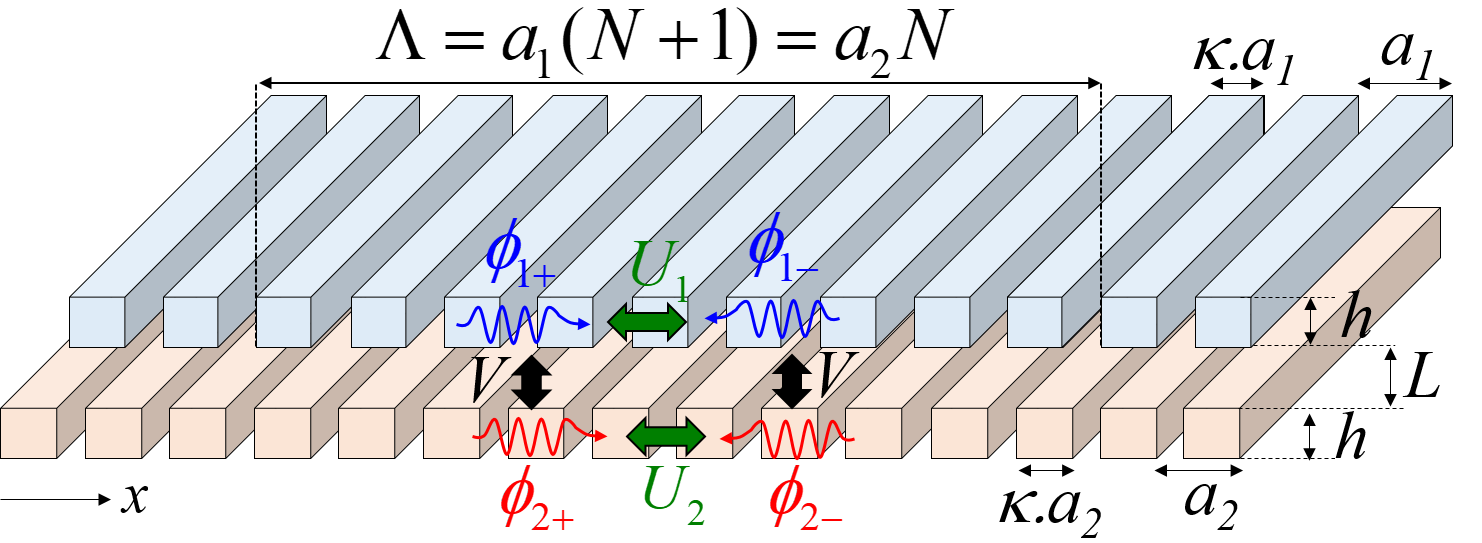}	\\
			\caption{Moir\'e superlattice of two gratings with of periods $a_1$ and $a_2$ satisfying $a_1/a_2=N/(N+1)$.}
			\label{fig:M}
		\end{center}
	\end{figure}
	
	Our system consists of two 1D photonic crystal slabs which are two subwavelength high refractive index contrast gratings (Fig \ref{fig:M}).  These gratings have the same subwavelength thickness $h$ and filling fraction $\kappa$ and are separated by only a subwavelength distance $L$. Their periods $a_1$ and $a_2$  are slightly different but satisfying the commensurate condition $a_1/a_2 =N/(N+1)$ for a natural number $N$. The period of the superlattice is given by $ \Lambda=(N+1)a_1=Na_2$, consisting of $N+1$ periods of the upper grating and $N$ period of the lower one. In the regime of $N\gg1$, a \textit{semi-continuous approach} can be implemented: the two gratings are almost identical and the \textit{moir\'e pattern} corresponds to a continuous shifting function $\delta(x)$ of the upper grating with respect to the lower grating, given by $\delta(0\leq x \leq \Lambda)=x/N$. The shifting $\delta$ sweeps an amount $a_0=(a_1+a_2)/2$ when $x$ varies across a moir\'e period. In other word, the moir\'e superlattice is obtained from the bilayer lattice by introducing a slight period mismatch: the period of the upper grating is shrunken from $a_0$ to $a_1$ and the period of the lower one is stretched from $a_0$ to $a_2$. This configuration leads to a modulated relative displacement $\delta(x)$ with respect to the coordinate $x$. Two special configurations of $\delta/a_0=0$ and 0.5 are referred to as $AA$- and $AB$-stackings, resembling the terminology in Bilayer Graphene structure~\cite{Rozhkov2016}. The moir\'e pattern is a period of a superlattice made of bilayer structures varying continuously from AA stacking to AB stacking.  The period mismatch leads to a Brillouin zone mismatch and the size of the mini Brillouin zone $K_M$ is given by $K_M= K_1-K_2$, where $K_1=2\pi/a_1$ and $K_2=2\pi/a_2$. \\
	
	In our perturbation approach, the dispersion characteristic of the  moir\'e superlattice is derived from two coupling mechanisms among forward $(\phi_{1+},\phi_{2+})$ and backward $(\phi_{1-},\phi_{2-})$ fundamental guided waves of the two noncorrugated slabs with effective refractive index: \textit{i) Intra-layer coupling} due to the diffractive processes\cite{Okamoto2006_CMT} between counter-propagating waves from the same layer. \textit{ii) Inter-layer coupling} via evanescence between co-propagating waves from separated layers. Using $(\phi_{1+},\phi_{1-},\phi_{2+},\phi_{2-})$ as basis, eigenmodes of the system are described by the following Hamiltonian (detailed derivation is given in the Supplemental Material): 
   \begin{small}
		\begin{equation}
		\label{eq:Hm}
		H=\left(\begin{matrix}
-iv\partial_x + \omega_1  & U_1 & V  & 0 \\
U_1 & iv\partial_x + \omega_1 & 0 & V  \\
V  & 0 & -iv\partial_x + \omega_2 & U_2e^{-i K_M x} \\
0 & V  & U_2e^{i K_M x} & iv\partial_x + \omega_2
\end{matrix}\right)
		\end{equation}
	\end{small}
	
	Here $U_{1,2}$ are the intra-layer coupling rates and $V$ is the inter-layer one; $v$ and $\omega_{1,2}$ are the group velocity and  offset energy of the guided waves at the Brillouin zone edge for each grating. A slight difference of values of the offset pulsation and the intra-layer coupling strength for each grating are due to the period mismatch, with $\omega_1\approx\omega_2\approx\omega_0$ and $U_1\approx U_2\approx U$ where $\omega_0$ and $U$ are the offset pulsation and the intra-layer coupling strength in the grating of period $a_0$. \\
	\begin{figure*}[hbt!]
		\begin{center}
			\includegraphics[width=0.75 \textwidth]{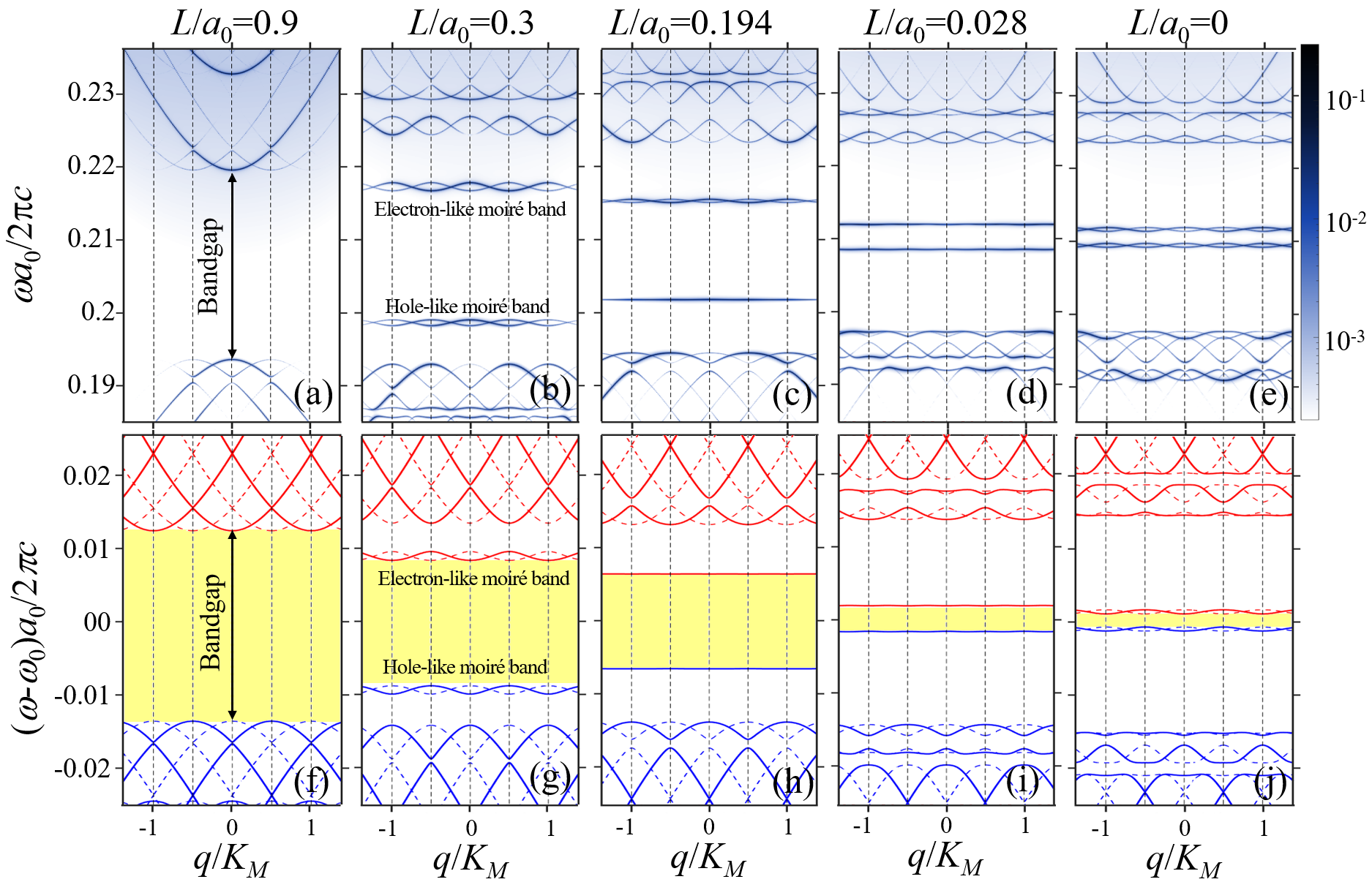}	
			\caption{(a-e) Simulated band structures corresponding to different $L$ values. The design for the simulation uses silicon ($n=3.54$) as the grating material, with $h=180\,nm$, $\kappa=0.8$, $a_0=(a_1+a_2)/2=300\,nm$ and  $N=13$. The photonic modes of uncoupled gratings are located below the light-line and the inter-layer coupling mechanisms, if not strong enough, would not be able to make these modes accessible for RCWA simulations. To solve this, a double period perturbation of $5\%$ is implemented for the design of each grating. The unit-cell in RCWA simulation consists of two moir\'e patterns: One is shrunken to $0.95\Lambda$, and the other one is dilated to $1.05\Lambda$. (f-i)  Calculations using the effective Hamiltonian of band structures shown in (a-e). To compare with the RCWA simulations having double period perturbation, dash-lines have been added, indicating the folding of the band structure. }
			\label{fig:band}
		\end{center}
	\end{figure*}
	The energy-momentum dispersion is simulated numerically using Rigorous Coupled-Wave Analysis (RCWA) method~\cite{Moharam:86,LIU20122233,Alonso-Alvarez2018}. The numerical results corresponding to $N=13$ when increasing the separation distance $L$ are presented in figures \ref{fig:band}a-e. When $L$ is comparable to $a_0$, the band structure is simply the folding of single layer dispersions (Fig \ref{fig:band}a). It suggests that the inter-layer coupling mechanism is negligible with respect to the intra-layer ones (i.e. $V \ll U$) for $L \gtrsim a_0$. In this configuration, a bandgap, purely due to the intra-layer coupling mechanism, is observed (Fig \ref{fig:band}a). In analogy to semiconductor terminology, we refer to these upper/lower bands as \textit{conduction-like/valence-like}. When $L\lesssim a_0$, the band hybridization due to the inter-layer coupling results in the formation of a pair of particle-hole minibands, referred to as electron-like/hole-like moir\'e band (Figs \ref{fig:band}b-e). These two bands emerge within the bandgap of uncoupled layers and are well isolated from the conduction/valence-like continuum. In the following, we will pay particular attention to the behavior of these two bands when tuning the inter-layer interaction.  One may note that with the choice of $a_0=300\,nm$, the spectral range of the these band is in the telecom (i.e.$\sim\,1.5\mu$m). Intriguingly, there are some specific values of $L$ at which the bandwidth of these bands becomes almost zero, and these moir\'e bands are nearly perfectly flat. Figures \ref{fig:band}c and \ref{fig:band}d depict the band structures with flat hole-like moir\'e band, and almost-flat electron-like band. Inspired by the analogy with the appearance of flatbands at magic angles in twisted bilayer graphene \cite{Tarnopolsky2019}, we called these values \textit{magic distances}. The moir\'e band structure is calculated using the Hamiltonian model given by Eq.~\eqref{eq:Hm}, taking $v$, $U$, $\omega_0$ and $V$ as input parameters. These parameters are retrieved from the simulation of single and bilayer lattice\cite{NHS2018,supp}. Figures \ref{fig:band}f-j depict the band structures obtained by analytical calculations. These results reproduce quantitatively the numerical results presented in Figs \ref{fig:band}a-e, showing the emergence of moir\'e states within the bandgap and their flattening at magic distances. Noticeably, there is a slight difference between simulation and analytical results:  the RCWA suggest that the flattening of the electron-like band always takes place at a slightly smaller distance $L$ than the one of the hole-like band, while the Hamiltonian model predicts that both bands become flat almost simultaneously. 
	
	\begin{figure}
		\begin{center}
			\includegraphics[width=0.4 \textwidth]{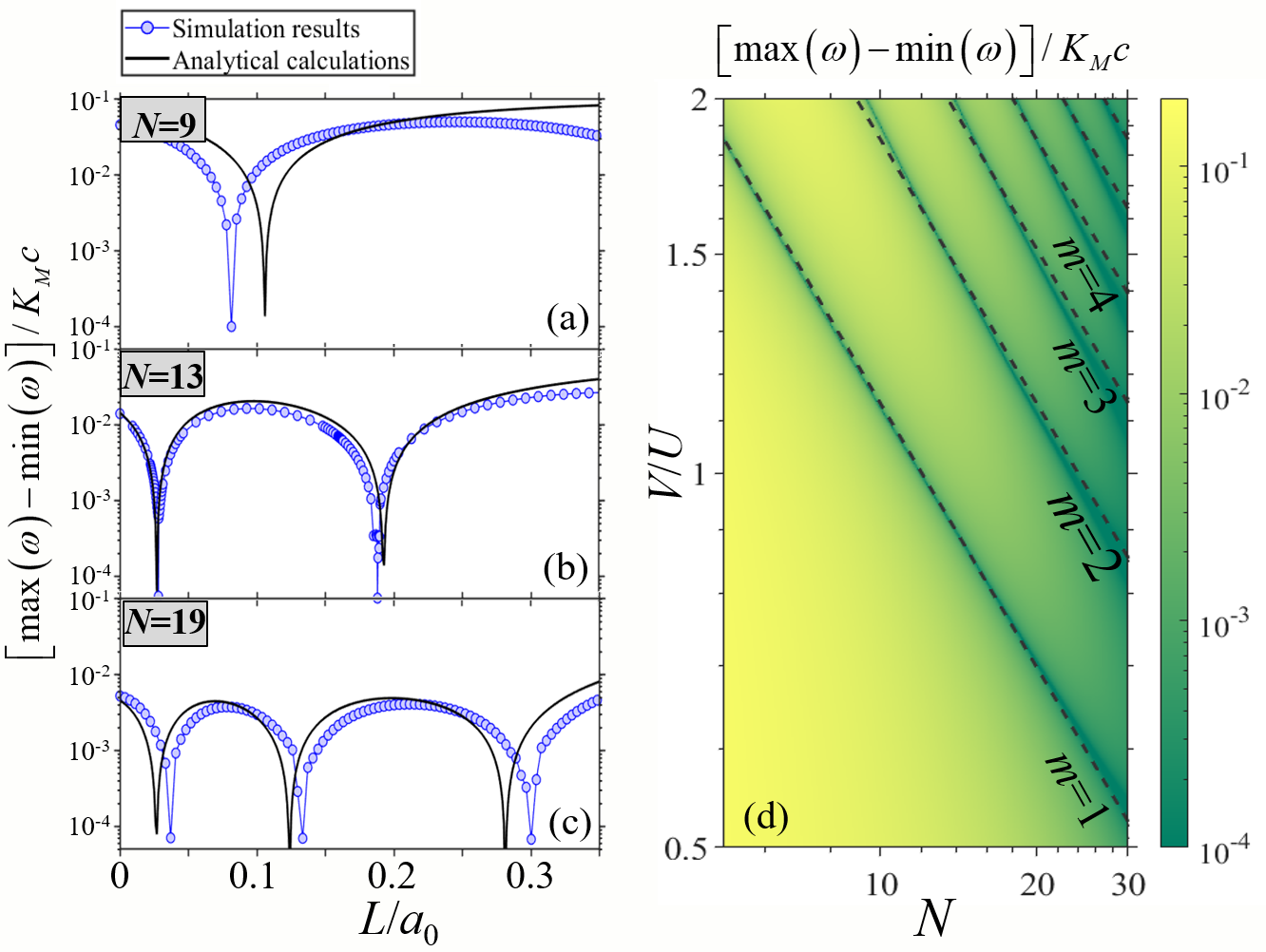}	
			\caption{(a-c) The global spectral bandwidth of the hole-like moir\'e band as a function of $L$ for different moir\'e patterns. Blue circles are results from RCWA simulations. Black lines are analytical calculations. (d) The global spectral bandwidth as a function of $V/U$ and $N$, with $U=U_0$. The dashed lines correspond to the empirical law ~\eqref{eq:magicNumber}. }
			\label{fig:magic}
		\end{center}
	\end{figure}

	The \textit{global spectral bandwidth}, defined as $\Delta\omega=\max_q(\omega) - \min_q(\omega)$, is used as the figure of merit to evaluate the flatness of moir\'e minibands. Figures \ref{fig:magic}a-c depict the global spectral bandwidth of the hole-like moir\'e band if different moir\'e superlattice ($N=9$, $13$ and $19$) when scanning $L$. These results confirm the existence of magic distances, corresponding to the bandwidth vanishings. All of the analytical calculations are obtained with the same set of parameters that are previously presented. We highlight that the Hamiltonian model provides almost perfectly both the number of magic distances and its values.
	
	For each moir\'e superlattice (i.e. a given $N$), our design exhibits two adjustable parameters: i) The distance $L$ for tuning the inter-layer coupling $V$ ($V=V_0$ when $L=0$ and decreasing exponentially when increasing $L$\cite{supp})  ; ii) The filling fraction $\kappa$, defined in Fig.\ref{fig:M}, for tuning the intra-layer coupling $U$ ($U=0$ when $\kappa=1$ and increasing when decreasing $\kappa$\cite{supp}). Up to now, we have been investigating flatband emergence by scanning $L$ while fixing $\kappa=0.8$ (i.e. $U=U_0$). However, the direct parameters of the Hamiltonian \eqref{eq:Hm} are $U$, $V$ and $N$ (from $K_M$). Thus a complete picture of magic configuration is captured when varying both $V/U$ (i.e. competition between inter versus intralayer coupling) and $N$ (i.e. moir\'e pattern). Figure~\ref{fig:magic}d presents the global bandwidth when scanning $N$ and $V/U$ within a reasonable range~\footnote{$N$ is varied from 5 to 30  (If $N$ is too small, the continuum model is not valid. If $N$ is too big, the Brillouin zone becomes too small and bands are naturally very flat). $V/U$ is varied from 0.5 to 2 (If $V/U$ is too small, the approximation $L\ll a_0$ is not valid. If $V/U >2$, the two moir\'e bands merge\cite{supp})}. The observed ``resonant dips'' correspond to different magic configurations. Dimensional analysis of Hamiltonian \eqref{eq:Hm} suggests that our system is driven by two dimensionless ratios  $V/U$, and $U/K_M \sim NU$\cite{supp}. Indeed,  fitting the resonances of Fig.\ref{fig:magic}d by a power law, we obtain a very simple empirical relation between there two dimensionless parameters:
	\begin{equation}\label{eq:magicNumber}
	N_mU = m  \times \eta \times \left(\frac{V}{U}\right)^{\gamma}\;\;,\;m=1,2,3...
	\end{equation}
	with the  $\gamma\approx-1.42$, $\eta\approx 12 U_0 $ , and $m$ is the \emph{counting order} of the  magic configuration. We note that $N$ is the ``moir\'e parameter'' in our system and playing the same role as the \textit{twist angle} in twisted bilayer graphene (each value of moir\'e parameter defines a moir\'e pattern)\cite{Santos2007,Santos2012}.
	Therefore, the good metric for magic configurations is the \textit{magic number} $N_m$, and Eq.~\eqref{eq:magicNumber} provides the design rule to achieve them. The analogy and similitude between this law and the one for magic angles in twisted bi-layer graphene~\cite{Tarnopolsky2019} are striking and  we expect an appealing interpretation for this simple relation.

	\begin{figure}[hbt!]
		\begin{center}
			\includegraphics[width=0.45 \textwidth]{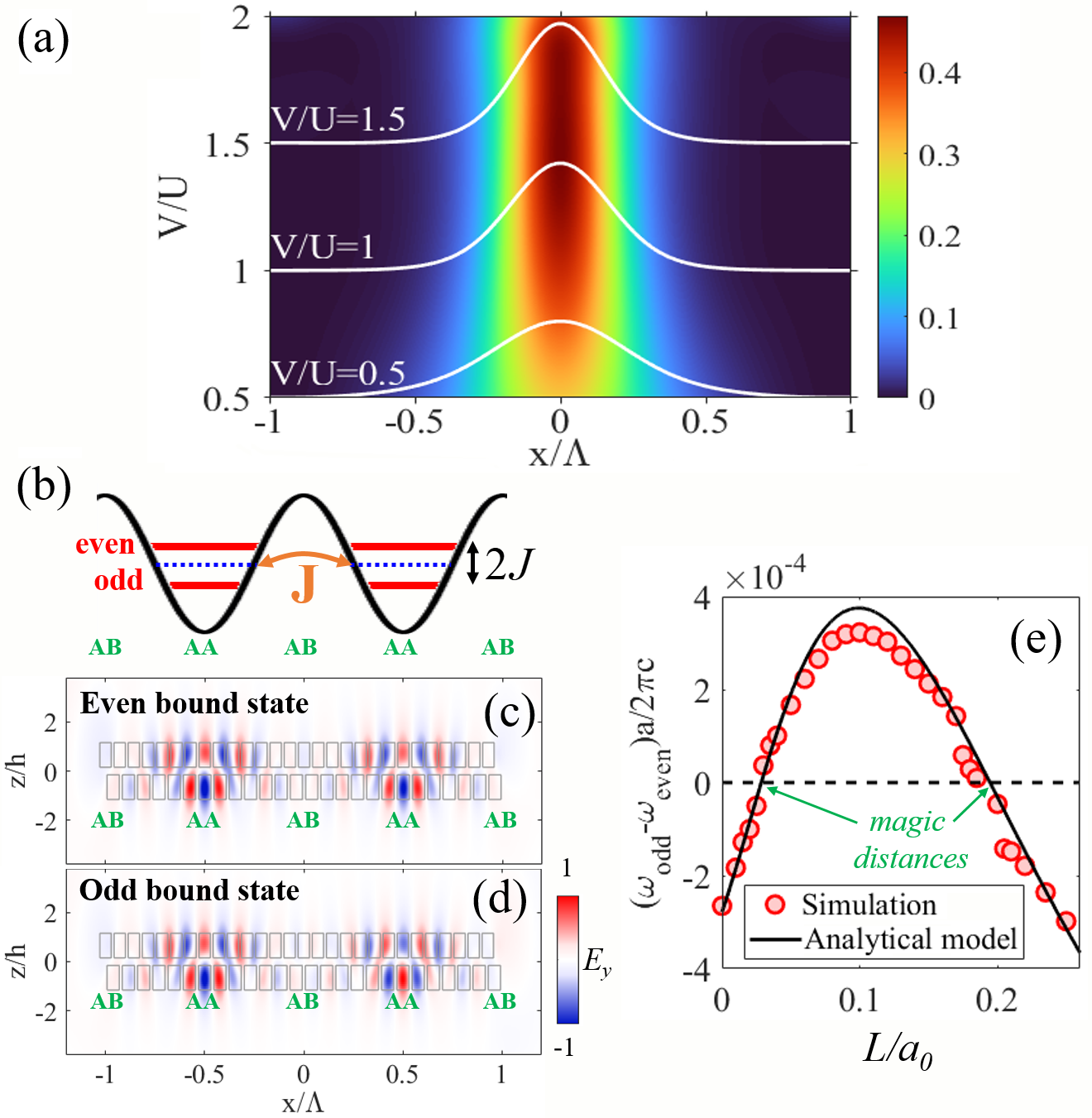}	
			\caption{(a) Wannier function, calculated by the twisted parallel transport gauge~\cite{Vanderbilt2018a}, of the hole-like moir\'e band when scanning the $V/U$ ratio. (b) Sketch of a hole-like ``diatomic molecule'' made of two moir\'e cells.  (c,d) The field distribution of the two hole-like bound states obtained by Finite-difference time-domain  (FDTD) simulations. The chosen moir\'e design is the same as the one in RCWA simulations in Fig.~\ref{fig:band}. (e) The energy splitting between the two bound states as a function of $L$. Red circles are results from FDTD simulations, and the solid black line is the result from the effective Hamiltonian. For the analytical calculation, the boundary condition is chosen so that outside of the moir\'e molecule is bilayer structure of $AB$ sites, and the bound states are calculated by the transfer matrix method~\cite{Davies1998a,Nguyen2008a}. For FDTD simulations, the structure only consists of two moir\'e cells.}
			\label{fig:Wannier_TightBinding}
		\end{center}
	\end{figure}

	
Knowing that flatband states would give rise to an unconventional localization regime~\cite{PRLmukher,PRLvincencio,Wang2020}, we now investigate the localization of light at magic configurations. A closer look at the two moir\'e bands in Figs.~\ref{fig:band} reveals that their dispersion characteristic are nearly single harmonic functions with the dominance of the first Fourier component with respect to higher-orders. Consequently, this suggests that each moir\'e band may be described by a textbook single-band tight-binding model with only a few nearest neighbour couplings taken into account. It is of interest to compute the Wannier functions for the band under consideration since Wannier functions are the natural basis for the tight-binding model~\cite{Mermin:Solid}.  Figure~\ref{fig:Wannier_TightBinding}(a) depicts the result of this calculation when scanning the ratio $V/U$, showing that more than 94\% of the Wannier density is located within a single moir\'e cell. Such a concentration confirms the use of this Wannier function as a pseudo-orbital wave function for the tight-binding model with nearest neighbour couplings. However, it is important to stress that the high concentration of the Wannier function is not necessarily related to flatband formations. Yet, the physics of the moir\'e bands can be captured quite well by a simple tight-binding scheme in the Wannier basis. In this scenario, the moir\'e superlattice engenders a periodic potential landscape with minima at AA sites. Trapped photons in the Wannier states can tunnel to the nearest neibour ones with tunnelling rate $J$ to form moir\'e bands of bandwidth $2\abs{J}$. As consequence, when the couple $\left(NU,V/U\right)$ satisfies the Eq.~\eqref{eq:magicNumber} of magic configurations, the only way to obtain dispersionless bands is that the effective tunneling rate $J$ becomes zero.  This leads to the tightly localization of light  within a single moir\'e cell at magic configurations. The compact localized states\cite{Maimaiti2017} of our localization is simply the Wannier function. We notice a resemblance of the flatband emergence in our system compared to the one in twisted bilayer graphene system~\cite{Tarnopolsky2019,Gadelha2021,nguyen2021electronic}: both correspond to the good localization at the AA sites.
	
	Keeping in mind the ability to localize light to a moir\'e period with high quality (albeit non-perfect), we investigate a much simpler problem: a ``diatomic molecule'' made of two moir\'e cells (Fig.\ref{fig:Wannier_TightBinding}b). Figures~\ref{fig:Wannier_TightBinding}c,d depicts the field distribution of the hole-like bound states with even (Fig.~\ref{fig:Wannier_TightBinding}c) and odd (Fig.~\ref{fig:Wannier_TightBinding}d) parity regarding the lateral mirror symmetry. The energy splitting when scanning the distance $L$ is presented in Fig.~\ref{fig:Wannier_TightBinding}e. Again, the results from the analytical model and numerical simulations show a very good agreement. Notably, these results demonstrate the crossing of these bound states exactly at the magic distances of the moir\'e superlattice from Fig.~\ref{fig:magic}b. Consequently, it supports that the tunnelling rate $J$ changes sign when scanning $L$ across a magic distance value and vanishes when $L$ takes a magic distance value.

	In conclusion, we have investigated theoretically the 1D moir\'e superlattice of bilayer photonic crystal.  All of analytical results derived from a simple effective Hamiltonian are in good agreement with numerical simulations, showing the emergence of flatband at magic configurations. The conditions for flatbands unify to a nontrivial relation between the counting order of the  magic configuration and the magic number, given by $N_m \sim m$. The physics of the moir\'e minibands is captured by a simple tight-binding model, resulting in localization of photonic states within a single moir\'e period at flatband configurations when the tunnelling rate  vanishes accidentally.  As fundamental perspective, the implementation of nonlinearity via Kerr nonlinearity~\cite{Rivas2020} or exciton-polariton platform~\cite{Goblot2019}, would pave the way to investigate the strongly correlated bosonic flatband physics~\cite{Leykam2017,Danieli2020,khalaf2021charged} with intriguing phases of 1D  matters \cite{giamarchi:2003}. For applications in optoelectronic devices, the  design in this work uses silicon as dielectric material, operating in the telecom range with feasible fabrication~\cite{Shuai2017,NHS2018,Cueff2019}, and is transferable to 1D integrated optics. The high sensitivity of dispersion band structure to the refractive index of surrounding medium (which determines the parameter $U$) and spacing medium (which determines the parameter $V$) can be harnessed for applications in sensing. Furthermore, the localization of light within the moir\'e period also suggests a unique way to engineer lattice of resonators of a high-quality factor for a phase-locked micro-laser array or high Purcell factor for tailoring spontaneous emission of quantum emitters. Another realization scheme is with dual-core fiber Bragg gratings~\cite{Mak98,Ahmed2019} to study soliton physics arising from photonic nonlinearity which will be greatly enhanced at flatband configurations\cite{Eggleton1996,Mak98,Ahmed2019}. 
	

	\textit{Acknowledgement:} The authors thank Stephen Carr, Nguyen Viet Hung,  and Steven H. Simon for fruitful discussions. The work is partly funded by the French National Research Agency (ANR) under the project POPEYE (ANR-17-CE24-0020) and the IDEXLYON from Université de Lyon, Scientific Breakthrough project TORE within the Programme Investissements d’Avenir (ANR-19-IDEX-0005).   DXN
	was supported by Brown Theoretical Physics Center.  
	HCN was supported by the Deutsche Forschungsgemeinschaft (DFG, German
	Research Foundation, project numbers 447948357 and 440958198), the Sino-German Center for Research Promotion (Project M-0294), and the ERC (Consolidator Grant 683107/TempoQ). RCWA simulations were performed on the CNRS/IN2P3 Computing Center in Lyon.
	
\bibliography{1DMoire.bib}

\begin{thebibliography}{46}%
\makeatletter
\providecommand \@ifxundefined [1]{%
 \@ifx{#1\undefined}
}%
\providecommand \@ifnum [1]{%
 \ifnum #1\expandafter \@firstoftwo
 \else \expandafter \@secondoftwo
 \fi
}%
\providecommand \@ifx [1]{%
 \ifx #1\expandafter \@firstoftwo
 \else \expandafter \@secondoftwo
 \fi
}%
\providecommand \natexlab [1]{#1}%
\providecommand \enquote  [1]{``#1''}%
\providecommand \bibnamefont  [1]{#1}%
\providecommand \bibfnamefont [1]{#1}%
\providecommand \citenamefont [1]{#1}%
\providecommand \href@noop [0]{\@secondoftwo}%
\providecommand \href [0]{\begingroup \@sanitize@url \@href}%
\providecommand \@href[1]{\@@startlink{#1}\@@href}%
\providecommand \@@href[1]{\endgroup#1\@@endlink}%
\providecommand \@sanitize@url [0]{\catcode `\\12\catcode `\$12\catcode
  `\&12\catcode `\#12\catcode `\^12\catcode `\_12\catcode `\%12\relax}%
\providecommand \@@startlink[1]{}%
\providecommand \@@endlink[0]{}%
\providecommand \url  [0]{\begingroup\@sanitize@url \@url }%
\providecommand \@url [1]{\endgroup\@href {#1}{\urlprefix }}%
\providecommand \urlprefix  [0]{URL }%
\providecommand \Eprint [0]{\href }%
\providecommand \doibase [0]{https://doi.org/}%
\providecommand \selectlanguage [0]{\@gobble}%
\providecommand \bibinfo  [0]{\@secondoftwo}%
\providecommand \bibfield  [0]{\@secondoftwo}%
\providecommand \translation [1]{[#1]}%
\providecommand \BibitemOpen [0]{}%
\providecommand \bibitemStop [0]{}%
\providecommand \bibitemNoStop [0]{.\EOS\space}%
\providecommand \EOS [0]{\spacefactor3000\relax}%
\providecommand \BibitemShut  [1]{\csname bibitem#1\endcsname}%
\let\auto@bib@innerbib\@empty
\bibitem [{\citenamefont {Bistritzer}\ and\ \citenamefont
  {MacDonald}(2011)}]{Bistritzer:2010}%
  \BibitemOpen
  \bibfield  {author} {\bibinfo {author} {\bibfnamefont {R.}~\bibnamefont
  {Bistritzer}}\ and\ \bibinfo {author} {\bibfnamefont {A.~H.}\ \bibnamefont
  {MacDonald}},\ }\bibfield  {title} {\bibinfo {title} {Moir{\'e} bands in
  twisted double-layer graphene},\ }\href
  {https://doi.org/10.1073/pnas.1108174108} {\bibfield  {journal} {\bibinfo
  {journal} {Proceedings of the National Academy of Sciences}\ }\textbf
  {\bibinfo {volume} {108}},\ \bibinfo {pages} {12233} (\bibinfo {year}
  {2011})},\ \Eprint
  {https://arxiv.org/abs/https://www.pnas.org/content/108/30/12233.full.pdf}
  {https://www.pnas.org/content/108/30/12233.full.pdf} \BibitemShut {NoStop}%
\bibitem [{\citenamefont {Tarnopolsky}\ \emph {et~al.}(2019)\citenamefont
  {Tarnopolsky}, \citenamefont {Kruchkov},\ and\ \citenamefont
  {Vishwanath}}]{Tarnopolsky2019}%
  \BibitemOpen
  \bibfield  {author} {\bibinfo {author} {\bibfnamefont {G.}~\bibnamefont
  {Tarnopolsky}}, \bibinfo {author} {\bibfnamefont {A.~J.}\ \bibnamefont
  {Kruchkov}},\ and\ \bibinfo {author} {\bibfnamefont {A.}~\bibnamefont
  {Vishwanath}},\ }\bibfield  {title} {\bibinfo {title} {{Origin of Magic
  Angles in Twisted Bilayer Graphene}},\ }\href
  {https://doi.org/10.1103/physrevlett.122.106405} {\bibfield  {journal}
  {\bibinfo  {journal} {Phys. Rev. Lett.}\ }\textbf {\bibinfo {volume} {122}},\
  \bibinfo {pages} {106405} (\bibinfo {year} {2019})}\BibitemShut {NoStop}%
\bibitem [{\citenamefont {Lisi}\ \emph {et~al.}(2021)\citenamefont {Lisi},
  \citenamefont {Lu}, \citenamefont {Benschop}, \citenamefont {de~Jong},
  \citenamefont {Stepanov}, \citenamefont {Duran}, \citenamefont {Margot},
  \citenamefont {Cucchi}, \citenamefont {Cappelli}, \citenamefont {Hunter},
  \citenamefont {Tamai}, \citenamefont {Kandyba}, \citenamefont {Giampietri},
  \citenamefont {Barinov}, \citenamefont {Jobst}, \citenamefont {Stalman},
  \citenamefont {Leeuwenhoek}, \citenamefont {Watanabe}, \citenamefont
  {Taniguchi}, \citenamefont {Rademaker}, \citenamefont {van~der Molen},
  \citenamefont {Allan}, \citenamefont {Efetov},\ and\ \citenamefont
  {Baumberger}}]{Lisi2021}%
  \BibitemOpen
  \bibfield  {author} {\bibinfo {author} {\bibfnamefont {S.}~\bibnamefont
  {Lisi}}, \bibinfo {author} {\bibfnamefont {X.}~\bibnamefont {Lu}}, \bibinfo
  {author} {\bibfnamefont {T.}~\bibnamefont {Benschop}}, \bibinfo {author}
  {\bibfnamefont {T.~A.}\ \bibnamefont {de~Jong}}, \bibinfo {author}
  {\bibfnamefont {P.}~\bibnamefont {Stepanov}}, \bibinfo {author}
  {\bibfnamefont {J.~R.}\ \bibnamefont {Duran}}, \bibinfo {author}
  {\bibfnamefont {F.}~\bibnamefont {Margot}}, \bibinfo {author} {\bibfnamefont
  {I.}~\bibnamefont {Cucchi}}, \bibinfo {author} {\bibfnamefont
  {E.}~\bibnamefont {Cappelli}}, \bibinfo {author} {\bibfnamefont
  {A.}~\bibnamefont {Hunter}}, \bibinfo {author} {\bibfnamefont
  {A.}~\bibnamefont {Tamai}}, \bibinfo {author} {\bibfnamefont
  {V.}~\bibnamefont {Kandyba}}, \bibinfo {author} {\bibfnamefont
  {A.}~\bibnamefont {Giampietri}}, \bibinfo {author} {\bibfnamefont
  {A.}~\bibnamefont {Barinov}}, \bibinfo {author} {\bibfnamefont
  {J.}~\bibnamefont {Jobst}}, \bibinfo {author} {\bibfnamefont
  {V.}~\bibnamefont {Stalman}}, \bibinfo {author} {\bibfnamefont
  {M.}~\bibnamefont {Leeuwenhoek}}, \bibinfo {author} {\bibfnamefont
  {K.}~\bibnamefont {Watanabe}}, \bibinfo {author} {\bibfnamefont
  {T.}~\bibnamefont {Taniguchi}}, \bibinfo {author} {\bibfnamefont
  {L.}~\bibnamefont {Rademaker}}, \bibinfo {author} {\bibfnamefont {S.~J.}\
  \bibnamefont {van~der Molen}}, \bibinfo {author} {\bibfnamefont {M.~P.}\
  \bibnamefont {Allan}}, \bibinfo {author} {\bibfnamefont {D.~K.}\ \bibnamefont
  {Efetov}},\ and\ \bibinfo {author} {\bibfnamefont {F.}~\bibnamefont
  {Baumberger}},\ }\bibfield  {title} {\bibinfo {title} {{Observation of flat
  bands in twisted bilayer graphene}},\ }\href
  {https://doi.org/10.1038/s41567-020-01041-x} {\bibfield  {journal} {\bibinfo
  {journal} {Nature Physics}\ }\textbf {\bibinfo {volume} {17}},\ \bibinfo
  {pages} {189} (\bibinfo {year} {2021})}\BibitemShut {NoStop}%
\bibitem [{\citenamefont {Cao}\ \emph {et~al.}(2018)\citenamefont {Cao},
  \citenamefont {Fatemi}, \citenamefont {Fang}, \citenamefont {Watanabe},
  \citenamefont {Taniguchi}, \citenamefont {Kaxiras},\ and\ \citenamefont
  {Jarillo-Herrero}}]{Cao2018}%
  \BibitemOpen
  \bibfield  {author} {\bibinfo {author} {\bibfnamefont {Y.}~\bibnamefont
  {Cao}}, \bibinfo {author} {\bibfnamefont {V.}~\bibnamefont {Fatemi}},
  \bibinfo {author} {\bibfnamefont {S.}~\bibnamefont {Fang}}, \bibinfo {author}
  {\bibfnamefont {K.}~\bibnamefont {Watanabe}}, \bibinfo {author}
  {\bibfnamefont {T.}~\bibnamefont {Taniguchi}}, \bibinfo {author}
  {\bibfnamefont {E.}~\bibnamefont {Kaxiras}},\ and\ \bibinfo {author}
  {\bibfnamefont {P.}~\bibnamefont {Jarillo-Herrero}},\ }\bibfield  {title}
  {\bibinfo {title} {{Unconventional superconductivity in magic-angle graphene
  superlattices}},\ }\href {https://doi.org/10.1038/nature26160} {\bibfield
  {journal} {\bibinfo  {journal} {Nature}\ }\textbf {\bibinfo {volume} {556}},\
  \bibinfo {pages} {43} (\bibinfo {year} {2018})}\BibitemShut {NoStop}%
\bibitem [{\citenamefont {Arora}\ \emph {et~al.}(2020)\citenamefont {Arora},
  \citenamefont {Polski}, \citenamefont {Zhang}, \citenamefont {Thomson},
  \citenamefont {Choi}, \citenamefont {Kim}, \citenamefont {Lin}, \citenamefont
  {Wilson}, \citenamefont {Xu}, \citenamefont {Chu}, \citenamefont {Watanabe},
  \citenamefont {Taniguchi}, \citenamefont {Alicea},\ and\ \citenamefont
  {Nadj-Perge}}]{Arora2020}%
  \BibitemOpen
  \bibfield  {author} {\bibinfo {author} {\bibfnamefont {H.~S.}\ \bibnamefont
  {Arora}}, \bibinfo {author} {\bibfnamefont {R.}~\bibnamefont {Polski}},
  \bibinfo {author} {\bibfnamefont {Y.}~\bibnamefont {Zhang}}, \bibinfo
  {author} {\bibfnamefont {A.}~\bibnamefont {Thomson}}, \bibinfo {author}
  {\bibfnamefont {Y.}~\bibnamefont {Choi}}, \bibinfo {author} {\bibfnamefont
  {H.}~\bibnamefont {Kim}}, \bibinfo {author} {\bibfnamefont {Z.}~\bibnamefont
  {Lin}}, \bibinfo {author} {\bibfnamefont {I.~Z.}\ \bibnamefont {Wilson}},
  \bibinfo {author} {\bibfnamefont {X.}~\bibnamefont {Xu}}, \bibinfo {author}
  {\bibfnamefont {J.-H.}\ \bibnamefont {Chu}}, \bibinfo {author} {\bibfnamefont
  {K.}~\bibnamefont {Watanabe}}, \bibinfo {author} {\bibfnamefont
  {T.}~\bibnamefont {Taniguchi}}, \bibinfo {author} {\bibfnamefont
  {J.}~\bibnamefont {Alicea}},\ and\ \bibinfo {author} {\bibfnamefont
  {S.}~\bibnamefont {Nadj-Perge}},\ }\bibfield  {title} {\bibinfo {title}
  {{Superconductivity in metallic twisted bilayer graphene stabilized by
  WSe2}},\ }\href {https://doi.org/10.1038/s41586-020-2473-8} {\bibfield
  {journal} {\bibinfo  {journal} {Nature}\ }\textbf {\bibinfo {volume} {583}},\
  \bibinfo {pages} {379} (\bibinfo {year} {2020})}\BibitemShut {NoStop}%
\bibitem [{\citenamefont {Stepanov}\ \emph {et~al.}(2020)\citenamefont
  {Stepanov}, \citenamefont {Das}, \citenamefont {Lu}, \citenamefont
  {Fahimniya}, \citenamefont {Watanabe}, \citenamefont {Taniguchi},
  \citenamefont {Koppens}, \citenamefont {Lischner}, \citenamefont {Levitov},\
  and\ \citenamefont {Efetov}}]{Stepanov2020}%
  \BibitemOpen
  \bibfield  {author} {\bibinfo {author} {\bibfnamefont {P.}~\bibnamefont
  {Stepanov}}, \bibinfo {author} {\bibfnamefont {I.}~\bibnamefont {Das}},
  \bibinfo {author} {\bibfnamefont {X.}~\bibnamefont {Lu}}, \bibinfo {author}
  {\bibfnamefont {A.}~\bibnamefont {Fahimniya}}, \bibinfo {author}
  {\bibfnamefont {K.}~\bibnamefont {Watanabe}}, \bibinfo {author}
  {\bibfnamefont {T.}~\bibnamefont {Taniguchi}}, \bibinfo {author}
  {\bibfnamefont {F.~H.~L.}\ \bibnamefont {Koppens}}, \bibinfo {author}
  {\bibfnamefont {J.}~\bibnamefont {Lischner}}, \bibinfo {author}
  {\bibfnamefont {L.}~\bibnamefont {Levitov}},\ and\ \bibinfo {author}
  {\bibfnamefont {D.~K.}\ \bibnamefont {Efetov}},\ }\bibfield  {title}
  {\bibinfo {title} {{Untying the insulating and superconducting orders in
  magic-angle graphene}},\ }\href {https://doi.org/10.1038/s41586-020-2459-6}
  {\bibfield  {journal} {\bibinfo  {journal} {Nature}\ }\textbf {\bibinfo
  {volume} {583}},\ \bibinfo {pages} {375} (\bibinfo {year}
  {2020})}\BibitemShut {NoStop}%
\bibitem [{\citenamefont {Song}\ \emph {et~al.}(2019)\citenamefont {Song},
  \citenamefont {Wang}, \citenamefont {Shi}, \citenamefont {Li}, \citenamefont
  {Fang},\ and\ \citenamefont {Bernevig}}]{Song2019}%
  \BibitemOpen
  \bibfield  {author} {\bibinfo {author} {\bibfnamefont {Z.}~\bibnamefont
  {Song}}, \bibinfo {author} {\bibfnamefont {Z.}~\bibnamefont {Wang}}, \bibinfo
  {author} {\bibfnamefont {W.}~\bibnamefont {Shi}}, \bibinfo {author}
  {\bibfnamefont {G.}~\bibnamefont {Li}}, \bibinfo {author} {\bibfnamefont
  {C.}~\bibnamefont {Fang}},\ and\ \bibinfo {author} {\bibfnamefont {B.~A.}\
  \bibnamefont {Bernevig}},\ }\bibfield  {title} {\bibinfo {title} {All magic
  angles in twisted bilayer graphene are topological},\ }\href
  {https://doi.org/10.1103/PhysRevLett.123.036401} {\bibfield  {journal}
  {\bibinfo  {journal} {Phys. Rev. Lett.}\ }\textbf {\bibinfo {volume} {123}},\
  \bibinfo {pages} {036401} (\bibinfo {year} {2019})}\BibitemShut {NoStop}%
\bibitem [{\citenamefont {Wu}\ \emph {et~al.}(2020)\citenamefont {Wu},
  \citenamefont {Zhang}, \citenamefont {Watanabe}, \citenamefont {Taniguchi},\
  and\ \citenamefont {Andrei}}]{wu2020chern}%
  \BibitemOpen
  \bibfield  {author} {\bibinfo {author} {\bibfnamefont {S.}~\bibnamefont
  {Wu}}, \bibinfo {author} {\bibfnamefont {Z.}~\bibnamefont {Zhang}}, \bibinfo
  {author} {\bibfnamefont {K.}~\bibnamefont {Watanabe}}, \bibinfo {author}
  {\bibfnamefont {T.}~\bibnamefont {Taniguchi}},\ and\ \bibinfo {author}
  {\bibfnamefont {E.~Y.}\ \bibnamefont {Andrei}},\ }\href@noop {} {\bibinfo
  {title} {Chern insulators and topological flat-bands in magic-angle twisted
  bilayer graphene}} (\bibinfo {year} {2020}),\ \Eprint
  {https://arxiv.org/abs/2007.03735} {arXiv:2007.03735 [cond-mat.mes-hall]}
  \BibitemShut {NoStop}%
\bibitem [{\citenamefont {Hu}\ \emph {et~al.}(2020{\natexlab{a}})\citenamefont
  {Hu}, \citenamefont {Ou}, \citenamefont {Si}, \citenamefont {Wu},
  \citenamefont {Wu}, \citenamefont {Dai}, \citenamefont {Krasnok},
  \citenamefont {Mazor}, \citenamefont {Zhang}, \citenamefont {Bao},
  \citenamefont {Qiu},\ and\ \citenamefont {Al{\`{u}}}}]{Hu2020a}%
  \BibitemOpen
  \bibfield  {author} {\bibinfo {author} {\bibfnamefont {G.}~\bibnamefont
  {Hu}}, \bibinfo {author} {\bibfnamefont {Q.}~\bibnamefont {Ou}}, \bibinfo
  {author} {\bibfnamefont {G.}~\bibnamefont {Si}}, \bibinfo {author}
  {\bibfnamefont {Y.}~\bibnamefont {Wu}}, \bibinfo {author} {\bibfnamefont
  {J.}~\bibnamefont {Wu}}, \bibinfo {author} {\bibfnamefont {Z.}~\bibnamefont
  {Dai}}, \bibinfo {author} {\bibfnamefont {A.}~\bibnamefont {Krasnok}},
  \bibinfo {author} {\bibfnamefont {Y.}~\bibnamefont {Mazor}}, \bibinfo
  {author} {\bibfnamefont {Q.}~\bibnamefont {Zhang}}, \bibinfo {author}
  {\bibfnamefont {Q.}~\bibnamefont {Bao}}, \bibinfo {author} {\bibfnamefont
  {C.~W.}\ \bibnamefont {Qiu}},\ and\ \bibinfo {author} {\bibfnamefont
  {A.}~\bibnamefont {Al{\`{u}}}},\ }\bibfield  {title} {\bibinfo {title}
  {{Topological polaritons and photonic magic angles in twisted $\alpha$-MoO3
  bilayers}},\ }\href {https://doi.org/10.1038/s41586-020-2359-9} {\bibfield
  {journal} {\bibinfo  {journal} {Nature}\ }\textbf {\bibinfo {volume} {582}},\
  \bibinfo {pages} {209} (\bibinfo {year} {2020}{\natexlab{a}})}\BibitemShut
  {NoStop}%
\bibitem [{\citenamefont {Hu}\ \emph {et~al.}(2020{\natexlab{b}})\citenamefont
  {Hu}, \citenamefont {Krasnok}, \citenamefont {Mazor}, \citenamefont {Qiu},\
  and\ \citenamefont {Al{\`{u}}}}]{Hu2020b}%
  \BibitemOpen
  \bibfield  {author} {\bibinfo {author} {\bibfnamefont {G.}~\bibnamefont
  {Hu}}, \bibinfo {author} {\bibfnamefont {A.}~\bibnamefont {Krasnok}},
  \bibinfo {author} {\bibfnamefont {Y.}~\bibnamefont {Mazor}}, \bibinfo
  {author} {\bibfnamefont {C.~W.}\ \bibnamefont {Qiu}},\ and\ \bibinfo {author}
  {\bibfnamefont {A.}~\bibnamefont {Al{\`{u}}}},\ }\bibfield  {title} {\bibinfo
  {title} {{Moir{\'{e}} Hyperbolic Metasurfaces}},\ }\href
  {https://doi.org/10.1021/acs.nanolett.9b05319} {\bibfield  {journal}
  {\bibinfo  {journal} {Nano Letters}\ }\textbf {\bibinfo {volume} {20}},\
  \bibinfo {pages} {3217} (\bibinfo {year} {2020}{\natexlab{b}})}\BibitemShut
  {NoStop}%
\bibitem [{\citenamefont {Wang}\ \emph {et~al.}(2020)\citenamefont {Wang},
  \citenamefont {Zheng}, \citenamefont {Chen}, \citenamefont {Huang},
  \citenamefont {Kartashov}, \citenamefont {Torner}, \citenamefont {Konotop},\
  and\ \citenamefont {Ye}}]{Wang2020}%
  \BibitemOpen
  \bibfield  {author} {\bibinfo {author} {\bibfnamefont {P.}~\bibnamefont
  {Wang}}, \bibinfo {author} {\bibfnamefont {Y.}~\bibnamefont {Zheng}},
  \bibinfo {author} {\bibfnamefont {X.}~\bibnamefont {Chen}}, \bibinfo {author}
  {\bibfnamefont {C.}~\bibnamefont {Huang}}, \bibinfo {author} {\bibfnamefont
  {Y.~V.}\ \bibnamefont {Kartashov}}, \bibinfo {author} {\bibfnamefont
  {L.}~\bibnamefont {Torner}}, \bibinfo {author} {\bibfnamefont {V.~V.}\
  \bibnamefont {Konotop}},\ and\ \bibinfo {author} {\bibfnamefont
  {F.}~\bibnamefont {Ye}},\ }\bibfield  {title} {\bibinfo {title}
  {{Localization and delocalization of light in photonic moir{\'{e}}
  lattices}},\ }\href {https://doi.org/10.1038/s41586-019-1851-6} {\bibfield
  {journal} {\bibinfo  {journal} {Nature}\ }\textbf {\bibinfo {volume} {577}},\
  \bibinfo {pages} {42} (\bibinfo {year} {2020})}\BibitemShut {NoStop}%
\bibitem [{\citenamefont {Lou}\ \emph {et~al.}(2021)\citenamefont {Lou},
  \citenamefont {Zhao}, \citenamefont {Minkov}, \citenamefont {Guo},
  \citenamefont {Orenstein},\ and\ \citenamefont {Fan}}]{Lou2021}%
  \BibitemOpen
  \bibfield  {author} {\bibinfo {author} {\bibfnamefont {B.}~\bibnamefont
  {Lou}}, \bibinfo {author} {\bibfnamefont {N.}~\bibnamefont {Zhao}}, \bibinfo
  {author} {\bibfnamefont {M.}~\bibnamefont {Minkov}}, \bibinfo {author}
  {\bibfnamefont {C.}~\bibnamefont {Guo}}, \bibinfo {author} {\bibfnamefont
  {M.}~\bibnamefont {Orenstein}},\ and\ \bibinfo {author} {\bibfnamefont
  {S.}~\bibnamefont {Fan}},\ }\bibfield  {title} {\bibinfo {title} {{Theory for
  Twisted Bilayer Photonic Crystal Slabs}},\ }\href
  {https://doi.org/10.1103/PhysRevLett.126.136101} {\bibfield  {journal}
  {\bibinfo  {journal} {Physical Review Letters}\ }\textbf {\bibinfo {volume}
  {126}},\ \bibinfo {pages} {136101} (\bibinfo {year} {2021})}\BibitemShut
  {NoStop}%
\bibitem [{\citenamefont {Dong}\ \emph {et~al.}(2021)\citenamefont {Dong},
  \citenamefont {Zhang}, \citenamefont {Li}, \citenamefont {Wang},
  \citenamefont {Yang}, \citenamefont {Rho}, \citenamefont {Wang},
  \citenamefont {Grigoropoulos}, \citenamefont {Wu},\ and\ \citenamefont
  {Yao}}]{Dong2021}%
  \BibitemOpen
  \bibfield  {author} {\bibinfo {author} {\bibfnamefont {K.}~\bibnamefont
  {Dong}}, \bibinfo {author} {\bibfnamefont {T.}~\bibnamefont {Zhang}},
  \bibinfo {author} {\bibfnamefont {J.}~\bibnamefont {Li}}, \bibinfo {author}
  {\bibfnamefont {Q.}~\bibnamefont {Wang}}, \bibinfo {author} {\bibfnamefont
  {F.}~\bibnamefont {Yang}}, \bibinfo {author} {\bibfnamefont {Y.}~\bibnamefont
  {Rho}}, \bibinfo {author} {\bibfnamefont {D.}~\bibnamefont {Wang}}, \bibinfo
  {author} {\bibfnamefont {C.~P.}\ \bibnamefont {Grigoropoulos}}, \bibinfo
  {author} {\bibfnamefont {J.}~\bibnamefont {Wu}},\ and\ \bibinfo {author}
  {\bibfnamefont {J.}~\bibnamefont {Yao}},\ }\bibfield  {title} {\bibinfo
  {title} {Flat bands in magic-angle bilayer photonic crystals at small
  twists},\ }\href {https://doi.org/10.1103/PhysRevLett.126.223601} {\bibfield
  {journal} {\bibinfo  {journal} {Phys. Rev. Lett.}\ }\textbf {\bibinfo
  {volume} {126}},\ \bibinfo {pages} {223601} (\bibinfo {year}
  {2021})}\BibitemShut {NoStop}%
\bibitem [{\citenamefont {Rozhkov}\ \emph {et~al.}(2016)\citenamefont
  {Rozhkov}, \citenamefont {Sboychakov}, \citenamefont {Rakhmanov},\ and\
  \citenamefont {Nori}}]{Rozhkov2016}%
  \BibitemOpen
  \bibfield  {author} {\bibinfo {author} {\bibfnamefont {A.~V.}\ \bibnamefont
  {Rozhkov}}, \bibinfo {author} {\bibfnamefont {A.~O.}\ \bibnamefont
  {Sboychakov}}, \bibinfo {author} {\bibfnamefont {A.~L.}\ \bibnamefont
  {Rakhmanov}},\ and\ \bibinfo {author} {\bibfnamefont {F.}~\bibnamefont
  {Nori}},\ }\bibfield  {title} {\bibinfo {title} {{Electronic properties of
  graphene-based bilayer systems}},\ }\href
  {https://doi.org/10.1016/j.physrep.2016.07.003} {\bibfield  {journal}
  {\bibinfo  {journal} {Physics Reports}\ }\textbf {\bibinfo {volume} {648}},\
  \bibinfo {pages} {1} (\bibinfo {year} {2016})},\ \Eprint
  {https://arxiv.org/abs/1511.06706} {arXiv:1511.06706} \BibitemShut {NoStop}%
\bibitem [{\citenamefont {Okamoto}(2006)}]{Okamoto2006_CMT}%
  \BibitemOpen
  \bibfield  {author} {\bibinfo {author} {\bibfnamefont {K.}~\bibnamefont
  {Okamoto}},\ }\bibfield  {title} {\bibinfo {title} {Chapter 4 - coupled mode
  theory},\ }in\ \href
  {https://doi.org/https://doi.org/10.1016/B978-012525096-2/50005-2} {\emph
  {\bibinfo {booktitle} {Fundamentals of Optical Waveguides (Second
  Edition)}}},\ \bibinfo {editor} {edited by\ \bibinfo {editor} {\bibfnamefont
  {K.}~\bibnamefont {Okamoto}}}\ (\bibinfo  {publisher} {Academic Press},\
  \bibinfo {address} {Burlington},\ \bibinfo {year} {2006})\ \bibinfo {edition}
  {second edition}\ ed.,\ pp.\ \bibinfo {pages} {159--207}\BibitemShut
  {NoStop}%
\bibitem [{\citenamefont {Moharam}\ and\ \citenamefont
  {Gaylord}(1986)}]{Moharam:86}%
  \BibitemOpen
  \bibfield  {author} {\bibinfo {author} {\bibfnamefont {M.~G.}\ \bibnamefont
  {Moharam}}\ and\ \bibinfo {author} {\bibfnamefont {T.~K.}\ \bibnamefont
  {Gaylord}},\ }\bibfield  {title} {\bibinfo {title} {Rigorous coupled-wave
  analysis of metallic surface-relief gratings},\ }\href
  {https://doi.org/10.1364/JOSAA.3.001780} {\bibfield  {journal} {\bibinfo
  {journal} {J. Opt. Soc. Am. A}\ }\textbf {\bibinfo {volume} {3}},\ \bibinfo
  {pages} {1780} (\bibinfo {year} {1986})}\BibitemShut {NoStop}%
\bibitem [{\citenamefont {Liu}\ and\ \citenamefont {Fan}(2012)}]{LIU20122233}%
  \BibitemOpen
  \bibfield  {author} {\bibinfo {author} {\bibfnamefont {V.}~\bibnamefont
  {Liu}}\ and\ \bibinfo {author} {\bibfnamefont {S.}~\bibnamefont {Fan}},\
  }\bibfield  {title} {\bibinfo {title} {S4 : A free electromagnetic solver for
  layered periodic structures},\ }\href
  {https://doi.org/https://doi.org/10.1016/j.cpc.2012.04.026} {\bibfield
  {journal} {\bibinfo  {journal} {Computer Physics Communications}\ }\textbf
  {\bibinfo {volume} {183}},\ \bibinfo {pages} {2233} (\bibinfo {year}
  {2012})}\BibitemShut {NoStop}%
\bibitem [{\citenamefont {Alonso-{\'A}lvarez}\ \emph
  {et~al.}(2018)\citenamefont {Alonso-{\'A}lvarez}, \citenamefont {Wilson},
  \citenamefont {Pearce}, \citenamefont {F{\"u}hrer}, \citenamefont {Farrell},\
  and\ \citenamefont {Ekins-Daukes}}]{Alonso-Alvarez2018}%
  \BibitemOpen
  \bibfield  {author} {\bibinfo {author} {\bibfnamefont {D.}~\bibnamefont
  {Alonso-{\'A}lvarez}}, \bibinfo {author} {\bibfnamefont {T.}~\bibnamefont
  {Wilson}}, \bibinfo {author} {\bibfnamefont {P.}~\bibnamefont {Pearce}},
  \bibinfo {author} {\bibfnamefont {M.}~\bibnamefont {F{\"u}hrer}}, \bibinfo
  {author} {\bibfnamefont {D.}~\bibnamefont {Farrell}},\ and\ \bibinfo {author}
  {\bibfnamefont {N.}~\bibnamefont {Ekins-Daukes}},\ }\bibfield  {title}
  {\bibinfo {title} {Solcore: a multi-scale, python-based library for modelling
  solar cells and semiconductor materials},\ }\href
  {https://doi.org/10.1007/s10825-018-1171-3} {\bibfield  {journal} {\bibinfo
  {journal} {Journal of Computational Electronics}\ }\textbf {\bibinfo {volume}
  {17}},\ \bibinfo {pages} {1099} (\bibinfo {year} {2018})}\BibitemShut
  {NoStop}%
\bibitem [{\citenamefont {Nguyen}\ \emph {et~al.}(2018)\citenamefont {Nguyen},
  \citenamefont {Dubois}, \citenamefont {Deschamps}, \citenamefont {Cueff},
  \citenamefont {Pardon}, \citenamefont {Leclercq}, \citenamefont {Seassal},
  \citenamefont {Letartre},\ and\ \citenamefont {Viktorovitch}}]{NHS2018}%
  \BibitemOpen
  \bibfield  {author} {\bibinfo {author} {\bibfnamefont {H.}~\bibnamefont
  {Nguyen}}, \bibinfo {author} {\bibfnamefont {F.}~\bibnamefont {Dubois}},
  \bibinfo {author} {\bibfnamefont {T.}~\bibnamefont {Deschamps}}, \bibinfo
  {author} {\bibfnamefont {S.}~\bibnamefont {Cueff}}, \bibinfo {author}
  {\bibfnamefont {A.}~\bibnamefont {Pardon}}, \bibinfo {author} {\bibfnamefont
  {J.-L.}\ \bibnamefont {Leclercq}}, \bibinfo {author} {\bibfnamefont
  {C.}~\bibnamefont {Seassal}}, \bibinfo {author} {\bibfnamefont
  {X.}~\bibnamefont {Letartre}},\ and\ \bibinfo {author} {\bibfnamefont
  {P.}~\bibnamefont {Viktorovitch}},\ }\bibfield  {title} {\bibinfo {title}
  {Symmetry breaking in photonic crystals: On-demand dispersion from flatband
  to dirac cones},\ }\bibfield  {journal} {\bibinfo  {journal} {Physical Review
  Letters}\ }\textbf {\bibinfo {volume} {120}},\ \href
  {https://doi.org/10.1103/physrevlett.120.066102}
  {10.1103/physrevlett.120.066102} (\bibinfo {year} {2018})\BibitemShut
  {NoStop}%
\bibitem [{sup()}]{supp}%
  \BibitemOpen
  \bibinfo {note} {See Supplemental Materials at (link) for full derivation
  details of the Hamiltonian models, the numerical simulations and parameter
  retrievals from band structure of single layer and bilayer lattices, as well
  as other further details.}\BibitemShut {Stop}%
\bibitem [{Note1()}]{Note1}%
  \BibitemOpen
  \bibinfo {note} {$N$ is varied from 5 to 30 (If $N$ is too small, the
  continuum model is not valid. If $N$ is too big, the Brillouin zone becomes
  too small and bands are naturally very flat). $V/U$ is varied from 0.5 to 2
  (If $V/U$ is too small, the approximation $L\ll a_0$ is not valid. If $V/U
  >2$, the two moir\'e bands merge\cite {supp})}\BibitemShut {NoStop}%
\bibitem [{\citenamefont {Lopes~dos Santos}\ \emph {et~al.}(2007)\citenamefont
  {Lopes~dos Santos}, \citenamefont {Peres},\ and\ \citenamefont
  {Castro~Neto}}]{Santos2007}%
  \BibitemOpen
  \bibfield  {author} {\bibinfo {author} {\bibfnamefont {J.~M.~B.}\
  \bibnamefont {Lopes~dos Santos}}, \bibinfo {author} {\bibfnamefont
  {N.~M.~R.}\ \bibnamefont {Peres}},\ and\ \bibinfo {author} {\bibfnamefont
  {A.~H.}\ \bibnamefont {Castro~Neto}},\ }\bibfield  {title} {\bibinfo {title}
  {Graphene bilayer with a twist: Electronic structure},\ }\href
  {https://doi.org/10.1103/PhysRevLett.99.256802} {\bibfield  {journal}
  {\bibinfo  {journal} {Phys. Rev. Lett.}\ }\textbf {\bibinfo {volume} {99}},\
  \bibinfo {pages} {256802} (\bibinfo {year} {2007})}\BibitemShut {NoStop}%
\bibitem [{\citenamefont {Lopes~dos Santos}\ \emph {et~al.}(2012)\citenamefont
  {Lopes~dos Santos}, \citenamefont {Peres},\ and\ \citenamefont
  {Castro~Neto}}]{Santos2012}%
  \BibitemOpen
  \bibfield  {author} {\bibinfo {author} {\bibfnamefont {J.~M.~B.}\
  \bibnamefont {Lopes~dos Santos}}, \bibinfo {author} {\bibfnamefont
  {N.~M.~R.}\ \bibnamefont {Peres}},\ and\ \bibinfo {author} {\bibfnamefont
  {A.~H.}\ \bibnamefont {Castro~Neto}},\ }\bibfield  {title} {\bibinfo {title}
  {Continuum model of the twisted graphene bilayer},\ }\href
  {https://doi.org/10.1103/PhysRevB.86.155449} {\bibfield  {journal} {\bibinfo
  {journal} {Phys. Rev. B}\ }\textbf {\bibinfo {volume} {86}},\ \bibinfo
  {pages} {155449} (\bibinfo {year} {2012})}\BibitemShut {NoStop}%
\bibitem [{\citenamefont {Vanderbilt}(2018)}]{Vanderbilt2018a}%
  \BibitemOpen
  \bibfield  {author} {\bibinfo {author} {\bibfnamefont {D.}~\bibnamefont
  {Vanderbilt}},\ }\href@noop {} {\emph {\bibinfo {title} {Berry phases in
  electronic structure theory}}}\ (\bibinfo  {publisher} {Cambridge University
  Press},\ \bibinfo {year} {2018})\BibitemShut {NoStop}%
\bibitem [{\citenamefont {Davies}(1998)}]{Davies1998a}%
  \BibitemOpen
  \bibfield  {author} {\bibinfo {author} {\bibfnamefont {J.~H.}\ \bibnamefont
  {Davies}},\ }\href@noop {} {\emph {\bibinfo {title} {The physics of
  low-dimensional semiconductors: an introduction}}}\ (\bibinfo  {publisher}
  {Cambridge University Press},\ \bibinfo {year} {1998})\BibitemShut {NoStop}%
\bibitem [{\citenamefont {Nguyen}\ \emph {et~al.}(2009)\citenamefont {Nguyen},
  \citenamefont {Hoang},\ and\ \citenamefont {Nguyen}}]{Nguyen2008a}%
  \BibitemOpen
  \bibfield  {author} {\bibinfo {author} {\bibfnamefont {H.~C.}\ \bibnamefont
  {Nguyen}}, \bibinfo {author} {\bibfnamefont {M.~T.}\ \bibnamefont {Hoang}},\
  and\ \bibinfo {author} {\bibfnamefont {V.~L.}\ \bibnamefont {Nguyen}},\
  }\bibfield  {title} {\bibinfo {title} {Quasi-bound states induced by
  one-dimensional potentials in graphene},\ }\href
  {https://doi.org/10.1103/PhysRevB.79.035411} {\bibfield  {journal} {\bibinfo
  {journal} {Phys. Rev. B}\ }\textbf {\bibinfo {volume} {79}},\ \bibinfo
  {pages} {035411} (\bibinfo {year} {2009})}\BibitemShut {NoStop}%
\bibitem [{\citenamefont {Mukherjee}\ \emph {et~al.}(2015)\citenamefont
  {Mukherjee}, \citenamefont {Spracklen}, \citenamefont {Choudhury},
  \citenamefont {Goldman}, \citenamefont {\"Ohberg}, \citenamefont
  {Andersson},\ and\ \citenamefont {Thomson}}]{PRLmukher}%
  \BibitemOpen
  \bibfield  {author} {\bibinfo {author} {\bibfnamefont {S.}~\bibnamefont
  {Mukherjee}}, \bibinfo {author} {\bibfnamefont {A.}~\bibnamefont
  {Spracklen}}, \bibinfo {author} {\bibfnamefont {D.}~\bibnamefont
  {Choudhury}}, \bibinfo {author} {\bibfnamefont {N.}~\bibnamefont {Goldman}},
  \bibinfo {author} {\bibfnamefont {P.}~\bibnamefont {\"Ohberg}}, \bibinfo
  {author} {\bibfnamefont {E.}~\bibnamefont {Andersson}},\ and\ \bibinfo
  {author} {\bibfnamefont {R.~R.}\ \bibnamefont {Thomson}},\ }\bibfield
  {title} {\bibinfo {title} {Observation of a localized flat-band state in a
  photonic lieb lattice},\ }\href
  {https://doi.org/10.1103/PhysRevLett.114.245504} {\bibfield  {journal}
  {\bibinfo  {journal} {Phys. Rev. Lett.}\ }\textbf {\bibinfo {volume} {114}},\
  \bibinfo {pages} {245504} (\bibinfo {year} {2015})}\BibitemShut {NoStop}%
\bibitem [{\citenamefont {Vicencio}\ \emph {et~al.}(2015)\citenamefont
  {Vicencio}, \citenamefont {Cantillano}, \citenamefont {Morales-Inostroza},
  \citenamefont {Real}, \citenamefont {Mej\'{\i}a-Cort\'es}, \citenamefont
  {Weimann}, \citenamefont {Szameit},\ and\ \citenamefont
  {Molina}}]{PRLvincencio}%
  \BibitemOpen
  \bibfield  {author} {\bibinfo {author} {\bibfnamefont {R.~A.}\ \bibnamefont
  {Vicencio}}, \bibinfo {author} {\bibfnamefont {C.}~\bibnamefont
  {Cantillano}}, \bibinfo {author} {\bibfnamefont {L.}~\bibnamefont
  {Morales-Inostroza}}, \bibinfo {author} {\bibfnamefont {B.}~\bibnamefont
  {Real}}, \bibinfo {author} {\bibfnamefont {C.}~\bibnamefont
  {Mej\'{\i}a-Cort\'es}}, \bibinfo {author} {\bibfnamefont {S.}~\bibnamefont
  {Weimann}}, \bibinfo {author} {\bibfnamefont {A.}~\bibnamefont {Szameit}},\
  and\ \bibinfo {author} {\bibfnamefont {M.~I.}\ \bibnamefont {Molina}},\
  }\bibfield  {title} {\bibinfo {title} {Observation of localized states in
  lieb photonic lattices},\ }\href
  {https://doi.org/10.1103/PhysRevLett.114.245503} {\bibfield  {journal}
  {\bibinfo  {journal} {Phys. Rev. Lett.}\ }\textbf {\bibinfo {volume} {114}},\
  \bibinfo {pages} {245503} (\bibinfo {year} {2015})}\BibitemShut {NoStop}%
\bibitem [{\citenamefont {Ashcroft}\ and\ \citenamefont
  {Mermin}(1976)}]{Mermin:Solid}%
  \BibitemOpen
  \bibfield  {author} {\bibinfo {author} {\bibfnamefont {N.~W.}\ \bibnamefont
  {Ashcroft}}\ and\ \bibinfo {author} {\bibfnamefont {N.~D.}\ \bibnamefont
  {Mermin}},\ }\href@noop {} {\emph {\bibinfo {title} {{S}olid {S}tate
  {P}hysics}}}\ (\bibinfo  {publisher} {Holt-Saunders},\ \bibinfo {year}
  {1976})\BibitemShut {NoStop}%
\bibitem [{\citenamefont {Maimaiti}\ \emph {et~al.}(2017)\citenamefont
  {Maimaiti}, \citenamefont {Andreanov}, \citenamefont {Park}, \citenamefont
  {Gendelman},\ and\ \citenamefont {Flach}}]{Maimaiti2017}%
  \BibitemOpen
  \bibfield  {author} {\bibinfo {author} {\bibfnamefont {W.}~\bibnamefont
  {Maimaiti}}, \bibinfo {author} {\bibfnamefont {A.}~\bibnamefont {Andreanov}},
  \bibinfo {author} {\bibfnamefont {H.~C.}\ \bibnamefont {Park}}, \bibinfo
  {author} {\bibfnamefont {O.}~\bibnamefont {Gendelman}},\ and\ \bibinfo
  {author} {\bibfnamefont {S.}~\bibnamefont {Flach}},\ }\bibfield  {title}
  {\bibinfo {title} {Compact localized states and flat-band generators in one
  dimension},\ }\href {https://doi.org/10.1103/PhysRevB.95.115135} {\bibfield
  {journal} {\bibinfo  {journal} {Phys. Rev. B}\ }\textbf {\bibinfo {volume}
  {95}},\ \bibinfo {pages} {115135} (\bibinfo {year} {2017})}\BibitemShut
  {NoStop}%
\bibitem [{\citenamefont {Gadelha}\ \emph {et~al.}(2021)\citenamefont
  {Gadelha}, \citenamefont {Ohlberg}, \citenamefont {Rabelo}, \citenamefont
  {Neto}, \citenamefont {Vasconcelos}, \citenamefont {Campos}, \citenamefont
  {Lemos}, \citenamefont {Ornelas}, \citenamefont {Miranda}, \citenamefont
  {Nadas}, \citenamefont {Santana}, \citenamefont {Watanabe}, \citenamefont
  {Taniguchi}, \citenamefont {van Troeye}, \citenamefont {Lamparski},
  \citenamefont {Meunier}, \citenamefont {Nguyen}, \citenamefont {Paszko},
  \citenamefont {Charlier}, \citenamefont {Campos}, \citenamefont
  {Can{\c{c}}ado}, \citenamefont {Medeiros-Ribeiro},\ and\ \citenamefont
  {Jorio}}]{Gadelha2021}%
  \BibitemOpen
  \bibfield  {author} {\bibinfo {author} {\bibfnamefont {A.~C.}\ \bibnamefont
  {Gadelha}}, \bibinfo {author} {\bibfnamefont {D.~A.~A.}\ \bibnamefont
  {Ohlberg}}, \bibinfo {author} {\bibfnamefont {C.}~\bibnamefont {Rabelo}},
  \bibinfo {author} {\bibfnamefont {E.~G.~S.}\ \bibnamefont {Neto}}, \bibinfo
  {author} {\bibfnamefont {T.~L.}\ \bibnamefont {Vasconcelos}}, \bibinfo
  {author} {\bibfnamefont {J.~L.}\ \bibnamefont {Campos}}, \bibinfo {author}
  {\bibfnamefont {J.~S.}\ \bibnamefont {Lemos}}, \bibinfo {author}
  {\bibfnamefont {V.}~\bibnamefont {Ornelas}}, \bibinfo {author} {\bibfnamefont
  {D.}~\bibnamefont {Miranda}}, \bibinfo {author} {\bibfnamefont
  {R.}~\bibnamefont {Nadas}}, \bibinfo {author} {\bibfnamefont {F.~C.}\
  \bibnamefont {Santana}}, \bibinfo {author} {\bibfnamefont {K.}~\bibnamefont
  {Watanabe}}, \bibinfo {author} {\bibfnamefont {T.}~\bibnamefont {Taniguchi}},
  \bibinfo {author} {\bibfnamefont {B.}~\bibnamefont {van Troeye}}, \bibinfo
  {author} {\bibfnamefont {M.}~\bibnamefont {Lamparski}}, \bibinfo {author}
  {\bibfnamefont {V.}~\bibnamefont {Meunier}}, \bibinfo {author} {\bibfnamefont
  {V.-H.}\ \bibnamefont {Nguyen}}, \bibinfo {author} {\bibfnamefont
  {D.}~\bibnamefont {Paszko}}, \bibinfo {author} {\bibfnamefont {J.-C.}\
  \bibnamefont {Charlier}}, \bibinfo {author} {\bibfnamefont {L.~C.}\
  \bibnamefont {Campos}}, \bibinfo {author} {\bibfnamefont {L.~G.}\
  \bibnamefont {Can{\c{c}}ado}}, \bibinfo {author} {\bibfnamefont
  {G.}~\bibnamefont {Medeiros-Ribeiro}},\ and\ \bibinfo {author} {\bibfnamefont
  {A.}~\bibnamefont {Jorio}},\ }\bibfield  {title} {\bibinfo {title}
  {{Localization of lattice dynamics in low-angle twisted bilayer graphene}},\
  }\href {https://doi.org/10.1038/s41586-021-03252-5} {\bibfield  {journal}
  {\bibinfo  {journal} {Nature}\ }\textbf {\bibinfo {volume} {590}},\ \bibinfo
  {pages} {405} (\bibinfo {year} {2021})}\BibitemShut {NoStop}%
\bibitem [{\citenamefont {Nguyen}\ \emph {et~al.}(2021)\citenamefont {Nguyen},
  \citenamefont {Paszko}, \citenamefont {Lamparski}, \citenamefont {Troeye},
  \citenamefont {Meunier},\ and\ \citenamefont
  {Charlier}}]{nguyen2021electronic}%
  \BibitemOpen
  \bibfield  {author} {\bibinfo {author} {\bibfnamefont {V.~H.}\ \bibnamefont
  {Nguyen}}, \bibinfo {author} {\bibfnamefont {D.}~\bibnamefont {Paszko}},
  \bibinfo {author} {\bibfnamefont {M.}~\bibnamefont {Lamparski}}, \bibinfo
  {author} {\bibfnamefont {B.~V.}\ \bibnamefont {Troeye}}, \bibinfo {author}
  {\bibfnamefont {V.}~\bibnamefont {Meunier}},\ and\ \bibinfo {author}
  {\bibfnamefont {J.~C.}\ \bibnamefont {Charlier}},\ }\href@noop {} {\bibinfo
  {title} {Electronic localization in small-angle twisted bilayer graphene}}
  (\bibinfo {year} {2021}),\ \Eprint {https://arxiv.org/abs/2102.05376}
  {arXiv:2102.05376 [cond-mat.mes-hall]} \BibitemShut {NoStop}%
\bibitem [{\citenamefont {Rivas}\ and\ \citenamefont
  {Molina}(2020)}]{Rivas2020}%
  \BibitemOpen
  \bibfield  {author} {\bibinfo {author} {\bibfnamefont {D.}~\bibnamefont
  {Rivas}}\ and\ \bibinfo {author} {\bibfnamefont {M.~I.}\ \bibnamefont
  {Molina}},\ }\bibfield  {title} {\bibinfo {title} {{Seltrapping in flat band
  lattices with nonlinear disorder}},\ }\href
  {https://doi.org/10.1038/s41598-020-62079-8} {\bibfield  {journal} {\bibinfo
  {journal} {Scientific Reports}\ }\textbf {\bibinfo {volume} {10}},\ \bibinfo
  {pages} {5229} (\bibinfo {year} {2020})}\BibitemShut {NoStop}%
\bibitem [{\citenamefont {Goblot}\ \emph {et~al.}(2019)\citenamefont {Goblot},
  \citenamefont {Rauer}, \citenamefont {Vicentini}, \citenamefont {Le~Boit\'e},
  \citenamefont {Galopin}, \citenamefont {Lema\^{\i}tre}, \citenamefont
  {Le~Gratiet}, \citenamefont {Harouri}, \citenamefont {Sagnes}, \citenamefont
  {Ravets}, \citenamefont {Ciuti}, \citenamefont {Amo},\ and\ \citenamefont
  {Bloch}}]{Goblot2019}%
  \BibitemOpen
  \bibfield  {author} {\bibinfo {author} {\bibfnamefont {V.}~\bibnamefont
  {Goblot}}, \bibinfo {author} {\bibfnamefont {B.}~\bibnamefont {Rauer}},
  \bibinfo {author} {\bibfnamefont {F.}~\bibnamefont {Vicentini}}, \bibinfo
  {author} {\bibfnamefont {A.}~\bibnamefont {Le~Boit\'e}}, \bibinfo {author}
  {\bibfnamefont {E.}~\bibnamefont {Galopin}}, \bibinfo {author} {\bibfnamefont
  {A.}~\bibnamefont {Lema\^{\i}tre}}, \bibinfo {author} {\bibfnamefont
  {L.}~\bibnamefont {Le~Gratiet}}, \bibinfo {author} {\bibfnamefont
  {A.}~\bibnamefont {Harouri}}, \bibinfo {author} {\bibfnamefont
  {I.}~\bibnamefont {Sagnes}}, \bibinfo {author} {\bibfnamefont
  {S.}~\bibnamefont {Ravets}}, \bibinfo {author} {\bibfnamefont
  {C.}~\bibnamefont {Ciuti}}, \bibinfo {author} {\bibfnamefont
  {A.}~\bibnamefont {Amo}},\ and\ \bibinfo {author} {\bibfnamefont
  {J.}~\bibnamefont {Bloch}},\ }\bibfield  {title} {\bibinfo {title} {Nonlinear
  polariton fluids in a flatband reveal discrete gap solitons},\ }\href
  {https://doi.org/10.1103/PhysRevLett.123.113901} {\bibfield  {journal}
  {\bibinfo  {journal} {Phys. Rev. Lett.}\ }\textbf {\bibinfo {volume} {123}},\
  \bibinfo {pages} {113901} (\bibinfo {year} {2019})}\BibitemShut {NoStop}%
\bibitem [{\citenamefont {Leykam}\ \emph {et~al.}(2017)\citenamefont {Leykam},
  \citenamefont {Bodyfelt}, \citenamefont {Desyatnikov},\ and\ \citenamefont
  {Flach}}]{Leykam2017}%
  \BibitemOpen
  \bibfield  {author} {\bibinfo {author} {\bibfnamefont {D.}~\bibnamefont
  {Leykam}}, \bibinfo {author} {\bibfnamefont {J.~D.}\ \bibnamefont
  {Bodyfelt}}, \bibinfo {author} {\bibfnamefont {A.~S.}\ \bibnamefont
  {Desyatnikov}},\ and\ \bibinfo {author} {\bibfnamefont {S.}~\bibnamefont
  {Flach}},\ }\bibfield  {title} {\bibinfo {title} {{Localization of weakly
  disordered flat band states}},\ }\href
  {https://doi.org/10.1140/epjb/e2016-70551-2} {\bibfield  {journal} {\bibinfo
  {journal} {The European Physical Journal B}\ }\textbf {\bibinfo {volume}
  {90}},\ \bibinfo {pages} {1} (\bibinfo {year} {2017})}\BibitemShut {NoStop}%
\bibitem [{\citenamefont {Danieli}\ \emph {et~al.}(2020)\citenamefont
  {Danieli}, \citenamefont {Andreanov},\ and\ \citenamefont
  {Flach}}]{Danieli2020}%
  \BibitemOpen
  \bibfield  {author} {\bibinfo {author} {\bibfnamefont {C.}~\bibnamefont
  {Danieli}}, \bibinfo {author} {\bibfnamefont {A.}~\bibnamefont {Andreanov}},\
  and\ \bibinfo {author} {\bibfnamefont {S.}~\bibnamefont {Flach}},\ }\bibfield
   {title} {\bibinfo {title} {Many-body flatband localization},\ }\href
  {https://doi.org/10.1103/PhysRevB.102.041116} {\bibfield  {journal} {\bibinfo
   {journal} {Phys. Rev. B}\ }\textbf {\bibinfo {volume} {102}},\ \bibinfo
  {pages} {041116} (\bibinfo {year} {2020})}\BibitemShut {NoStop}%
\bibitem [{\citenamefont {Khalaf}\ \emph {et~al.}(2021)\citenamefont {Khalaf},
  \citenamefont {Chatterjee}, \citenamefont {Bultinck}, \citenamefont
  {Zaletel},\ and\ \citenamefont {Vishwanath}}]{khalaf2021charged}%
  \BibitemOpen
  \bibfield  {author} {\bibinfo {author} {\bibfnamefont {E.}~\bibnamefont
  {Khalaf}}, \bibinfo {author} {\bibfnamefont {S.}~\bibnamefont {Chatterjee}},
  \bibinfo {author} {\bibfnamefont {N.}~\bibnamefont {Bultinck}}, \bibinfo
  {author} {\bibfnamefont {M.~P.}\ \bibnamefont {Zaletel}},\ and\ \bibinfo
  {author} {\bibfnamefont {A.}~\bibnamefont {Vishwanath}},\ }\href@noop {}
  {\bibinfo {title} {Charged skyrmions and topological origin of
  superconductivity in magic angle graphene}} (\bibinfo {year} {2021}),\
  \Eprint {https://arxiv.org/abs/2004.00638} {arXiv:2004.00638
  [cond-mat.str-el]} \BibitemShut {NoStop}%
\bibitem [{\citenamefont {Giamarchi}(2003)}]{giamarchi:2003}%
  \BibitemOpen
  \bibfield  {author} {\bibinfo {author} {\bibfnamefont {T.}~\bibnamefont
  {Giamarchi}},\ }\href
  {https://doi.org/10.1093/acprof:oso/9780198525004.001.0001} {\emph {\bibinfo
  {title} {Quantum {Physics} in {One} {Dimension}}}}\ (\bibinfo  {publisher}
  {Oxford University Press},\ \bibinfo {year} {2003})\BibitemShut {NoStop}%
\bibitem [{\citenamefont {Shuai}\ \emph {et~al.}(2017)\citenamefont {Shuai},
  \citenamefont {Zhao}, \citenamefont {Liu}, \citenamefont {Stambaugh},
  \citenamefont {Lawall},\ and\ \citenamefont {Zhou}}]{Shuai2017}%
  \BibitemOpen
  \bibfield  {author} {\bibinfo {author} {\bibfnamefont {Y.}~\bibnamefont
  {Shuai}}, \bibinfo {author} {\bibfnamefont {D.}~\bibnamefont {Zhao}},
  \bibinfo {author} {\bibfnamefont {Y.}~\bibnamefont {Liu}}, \bibinfo {author}
  {\bibfnamefont {C.}~\bibnamefont {Stambaugh}}, \bibinfo {author}
  {\bibfnamefont {J.}~\bibnamefont {Lawall}},\ and\ \bibinfo {author}
  {\bibfnamefont {W.}~\bibnamefont {Zhou}},\ }\bibfield  {title} {\bibinfo
  {title} {Coupled bilayer photonic crystal slab electro-optic spatial light
  modulators},\ }\href@noop {} {\bibfield  {journal} {\bibinfo  {journal} {IEEE
  Photonics Journal}\ }\textbf {\bibinfo {volume} {9}},\ \bibinfo {pages} {1}
  (\bibinfo {year} {2017})}\BibitemShut {NoStop}%
\bibitem [{\citenamefont {Cueff}\ \emph {et~al.}(2019)\citenamefont {Cueff},
  \citenamefont {Dubois}, \citenamefont {Huang}, \citenamefont {Li},
  \citenamefont {Zia}, \citenamefont {Letartre}, \citenamefont {Viktorovitch},\
  and\ \citenamefont {Nguyen}}]{Cueff2019}%
  \BibitemOpen
  \bibfield  {author} {\bibinfo {author} {\bibfnamefont {S.}~\bibnamefont
  {Cueff}}, \bibinfo {author} {\bibfnamefont {F.}~\bibnamefont {Dubois}},
  \bibinfo {author} {\bibfnamefont {M.~S.~R.}\ \bibnamefont {Huang}}, \bibinfo
  {author} {\bibfnamefont {D.}~\bibnamefont {Li}}, \bibinfo {author}
  {\bibfnamefont {R.}~\bibnamefont {Zia}}, \bibinfo {author} {\bibfnamefont
  {X.}~\bibnamefont {Letartre}}, \bibinfo {author} {\bibfnamefont
  {P.}~\bibnamefont {Viktorovitch}},\ and\ \bibinfo {author} {\bibfnamefont
  {H.~S.}\ \bibnamefont {Nguyen}},\ }\bibfield  {title} {\bibinfo {title}
  {Tailoring the local density of optical states and directionality of light
  emission by symmetry breaking},\ }\href
  {https://doi.org/10.1109/JSTQE.2019.2902915} {\bibfield  {journal} {\bibinfo
  {journal} {IEEE Journal of Selected Topics in Quantum Electronics}\ }\textbf
  {\bibinfo {volume} {25}},\ \bibinfo {pages} {1} (\bibinfo {year}
  {2019})}\BibitemShut {NoStop}%
\bibitem [{\citenamefont {Mak}\ \emph {et~al.}(1998)\citenamefont {Mak},
  \citenamefont {Chu},\ and\ \citenamefont {Malomed}}]{Mak98}%
  \BibitemOpen
  \bibfield  {author} {\bibinfo {author} {\bibfnamefont {W.~C.~K.}\
  \bibnamefont {Mak}}, \bibinfo {author} {\bibfnamefont {P.~L.}\ \bibnamefont
  {Chu}},\ and\ \bibinfo {author} {\bibfnamefont {B.~A.}\ \bibnamefont
  {Malomed}},\ }\bibfield  {title} {\bibinfo {title} {Solitary waves in coupled
  nonlinear waveguides with bragg gratings},\ }\href
  {https://doi.org/10.1364/JOSAB.15.001685} {\bibfield  {journal} {\bibinfo
  {journal} {J. Opt. Soc. Am. B}\ }\textbf {\bibinfo {volume} {15}},\ \bibinfo
  {pages} {1685} (\bibinfo {year} {1998})}\BibitemShut {NoStop}%
\bibitem [{\citenamefont {Ahmed}\ and\ \citenamefont {Atai}(2019)}]{Ahmed2019}%
  \BibitemOpen
  \bibfield  {author} {\bibinfo {author} {\bibfnamefont {T.}~\bibnamefont
  {Ahmed}}\ and\ \bibinfo {author} {\bibfnamefont {J.}~\bibnamefont {Atai}},\
  }\bibfield  {title} {\bibinfo {title} {{Soliton-soliton dynamics in a
  dual-core system with separated nonlinearity and nonuniform Bragg grating}},\
  }\href {https://doi.org/10.1007/s11071-019-05069-4} {\bibfield  {journal}
  {\bibinfo  {journal} {Nonlinear Dynamics}\ }\textbf {\bibinfo {volume}
  {97}},\ \bibinfo {pages} {1515} (\bibinfo {year} {2019})}\BibitemShut
  {NoStop}%
\bibitem [{\citenamefont {Eggleton}\ \emph {et~al.}(1996)\citenamefont
  {Eggleton}, \citenamefont {Slusher}, \citenamefont {de~Sterke}, \citenamefont
  {Krug},\ and\ \citenamefont {Sipe}}]{Eggleton1996}%
  \BibitemOpen
  \bibfield  {author} {\bibinfo {author} {\bibfnamefont {B.~J.}\ \bibnamefont
  {Eggleton}}, \bibinfo {author} {\bibfnamefont {R.~E.}\ \bibnamefont
  {Slusher}}, \bibinfo {author} {\bibfnamefont {C.~M.}\ \bibnamefont
  {de~Sterke}}, \bibinfo {author} {\bibfnamefont {P.~A.}\ \bibnamefont
  {Krug}},\ and\ \bibinfo {author} {\bibfnamefont {J.~E.}\ \bibnamefont
  {Sipe}},\ }\bibfield  {title} {\bibinfo {title} {Bragg grating solitons},\
  }\href {https://doi.org/10.1103/PhysRevLett.76.1627} {\bibfield  {journal}
  {\bibinfo  {journal} {Phys. Rev. Lett.}\ }\textbf {\bibinfo {volume} {76}},\
  \bibinfo {pages} {1627} (\bibinfo {year} {1996})}\BibitemShut {NoStop}%
\bibitem [{Note2()}]{Note2}%
  \BibitemOpen
  \bibinfo {note} {Here, we ignore the coupling between the positive (negative)
  mode on the upper layer and the negative (positive) mode on the lower layer.
  This coupling includes a fast oscillation factor dues to the fact that the
  positive mode and the negative mode have different wave vectors.}\BibitemShut
  {Stop}%
\bibitem [{Note3()}]{Note3}%
  \BibitemOpen
  \bibinfo {note} {This can be demonstrated by using $\sigma _z \sigma _{\pm }
  \sigma _z^{\dagger } = -\sigma _{\pm }$ and $T_{\Lambda } e^{-i \theta \sigma
  _z} T_{\Lambda }^{\dagger } = -e^{-i \theta \sigma _z}$.}\BibitemShut {Stop}%
\bibitem [{\citenamefont {Rackauckas}\ and\ \citenamefont
  {Nie}(2017)}]{Rackauckas2017a}%
  \BibitemOpen
  \bibfield  {author} {\bibinfo {author} {\bibfnamefont {C.}~\bibnamefont
  {Rackauckas}}\ and\ \bibinfo {author} {\bibfnamefont {Q.}~\bibnamefont
  {Nie}},\ }\bibfield  {title} {\bibinfo {title} {Differentialequations.jl--a
  performant and feature-rich ecosystem for solving differential equations in
  julia},\ }\href@noop {} {\bibfield  {journal} {\bibinfo  {journal} {Journal
  of Open Research Software}\ }\textbf {\bibinfo {volume} {5}} (\bibinfo {year}
  {2017})}\BibitemShut {NoStop}%
\end{thebibliography}%


%
\clearpage
\newpage

\begin{widetext}
	\begin{center}
		\textbf{\large --- Supplementary Material ---\\ Magic configurations in Moir\'e Superlattice of Bilayer Photonic crystal: \\  Almost-Perfect Flatbands and Unconventional Localization}\\
		\medskip
		\text{Dung Xuan Nguyen, Xavier Letartre, Emmanuel Drouard, Pierre Viktorovitch, H Chau Nguyen, Hai Son Nguyen }
	\end{center}
	\setcounter{equation}{0}
	\setcounter{figure}{0}
	\setcounter{table}{0}
	\setcounter{page}{1}
	\makeatletter
	\renewcommand{\theequation}{S\arabic{equation}}
	\renewcommand{\thefigure}{S\arabic{figure}}
	\renewcommand{\bibnumfmt}[1]{[S#1]}
		
		
		\section{Ab initio derivation of Moir\'e lattice Hamiltonian}
		\label{sec:derivation}
		In this section, we provide the detailed derivation of the effective Hamiltonian in the main text. 
		\subsection{Hamiltonian of a single grating wave-guide}
		\subsubsection{Wave function a single grating wave-guide}
		In perturbation theory, the eigenmodes in photonic grating are constituted by the coupling between forward $\varphi_+(k\geq 0)$ and backward $\varphi_-(k\leq0)$ propagating waves of the non-corrugated waveguide of effective refractive index  (see Fig \ref{fig:dis1}a). Here the \textit{wave-function} $\varphi$ corresponds to the electric field $E_y$ for TE modes, and the magnetic field $H_y$ for TM modes.  The dispersion characteristic $\omega_+(k\geq0)$ and $\omega_-(k\leq0) $, $\omega_+(k) = \omega_-(-k)$, of these guided modes lies below the light-line (see Fig \ref{fig:dis1}b) and are obtained by solving Maxwell equations of planar waveguide with effective refractive index.
		We can extend the definition of  positive and negative wavefunctions for any $k$ value by replacing $\varphi_{\pm}(k)$ by $\mathbf{\Phi}_{\pm}(k)$, given by:
		\begin{equation}
			\label{eq:modesk}
			\mathbf{\Phi}_\pm(k)=\Theta(\pm k)\varphi_\pm\left(k\right),
		\end{equation}
		where $\Theta$ is the Heaviside function, $\theta(x)=1$ if  $x\geq0$ and $\theta (x) = 0$ if $ x<0$. With such definition, the spatial wave-function $\Phi_\pm(x)$ of positive and negative modes is obtained by the Fourier transform of $\Phi_\pm(k)$:
		\begin{equation}
			\mathbf{\Phi}_\pm(x)=\int \frac{\dd k}{2\pi}\,\mathbf{\Phi}_\pm(k)e^{ikx}. 
		\end{equation}
		With a spatial period $a$, the reciprocal lattice vector is given by $K_0=\frac{2\pi}{a_0}$. High symmetry points in the momentum space are at wavevectors $\frac{lK_0}{2}$ with $l \in\mathbb{Z}$. A given odd(even) value of $l$ corresponds to a $X(\Gamma)$ point of the  BZs. The \textit{effective} wave-functions of positive (negative mode) near the high symmetry point $\frac{lK_0}{2}$ (-$\frac{lK_0}{2}$) are defined by:
		\begin{equation}
			\label{eq:efwq}
			\Phi_{l,\pm}(q)=\Phi_{\pm}\left(\pm\frac{l K_0}{2}+q\right),\quad q\in\left[-\frac{K_0}{4},\frac{K_0}{4}\right],
		\end{equation}
		and
		\begin{equation}
			\label{eq:Fourier}
			\Phi_{l,\pm}(x)=\int_{-\frac{K_0}{4}}^{\frac{K_0}{4}}\frac{\dd q}{2\pi}\, \Phi_{l,\pm}(q)e^{iqx}.
		\end{equation}
		We verify easily the relation between $\Phi_\pm(x)$ and $\Phi_{l,\pm}(x)$, given by: 
		\begin{equation}
			\label{eq:efwx}
			\mathbf{\Phi}_\pm(x)=\sum_{l\in \mathbb{Z}}e^{\pm i\frac{lK_0}{2}x}\Phi_{l,\pm}(x)
		\end{equation}
		\begin{figure}[b]
			\begin{center}
				\includegraphics[width=0.7 \textwidth]{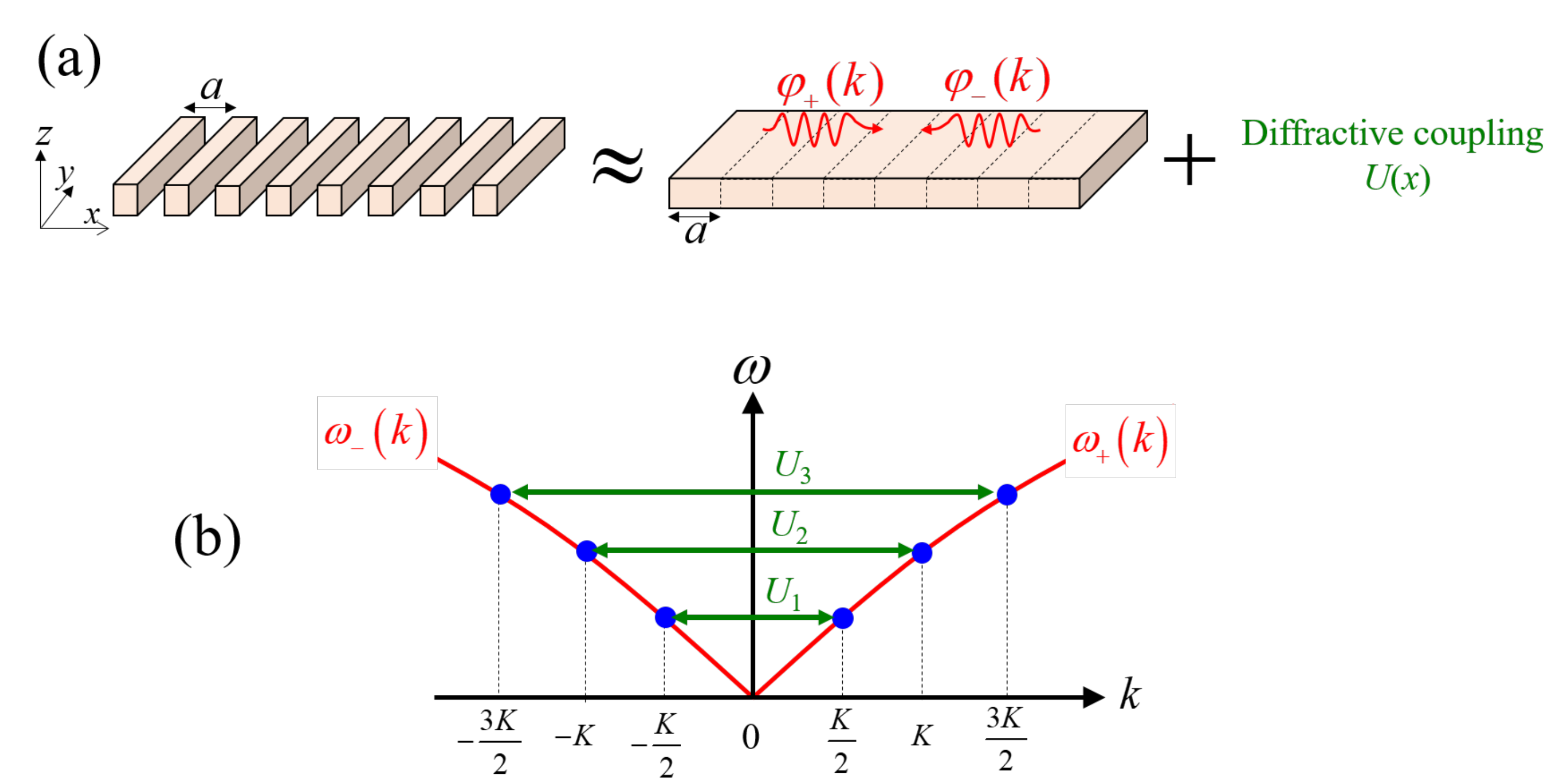}	
				\caption{(a) Sketch of the photonic grating and the non-corrugated waveguide in effective refractive index approach. (b) Dispersions relation of photonic guided modes $\omega_\pm(k)$ and the coupling between positive and negative modes due to periodic potentials $U_D(x)=\sum_{l}U_l e^{i \frac{2l\pi}{a}x}$.}
				\label{fig:dis1}
			\end{center}
		\end{figure}
		Since band structures are mostly studied in the vicinity of a high symmetry point of the BZs, the most appropriate basis in real space and momentum space given by:
		\begin{align}
			\label{eq:basisPsi_x}	
			\Psi_l(x)=\left(\begin{matrix}\Phi_{l,+}(x) \\ \Phi_{l,-}(x)\end{matrix}\right). 
		\end{align} 
		Note that due to the fact that positive mode has positive wave-vectors and the negative mode has negative wave-vectors, only $l \in \mathbb{N}^*$ appears in the definitions \eqref{eq:basisPsi_x}. 
		In the vicinity $|q|\ll \frac{K_0}{4}$ of high symmetry points (the blue points in Fig \ref{fig:dis1}b) in momentum space, these relations can be approximated by
		\begin{equation}
			\omega_+ \left(\frac{lK_0}{2}+q\right)
			\approx\omega_{0l} + v_l q, \quad
			\label{eq:dis_neg}
			\omega_- \left(-\frac{lK_0}{2}+q\right)
			\approx\omega_{0l} - v_l q 
		\end{equation}
		We have the \textit{effective} free Hamiltonian density in momentum space  $H^\text{free}_{(l)}(q)$ near the high symmetry points in the momentum space 
		\begin{equation}
			\label{eq:Hfreeq}
			\!\!\!\! H^\text{free}_{l}(q) = \left(\omega_{0l}+v_l q \right)\Phi_{l,+}^\dagger(q)\Phi_{l,+}(q) + \left(\omega_{0l}-v_l q \right)\Phi_{l,-}^\dagger(q)\Phi_{l,-}(q),   
		\end{equation}
		\subsubsection{Diffractive coupling between counter-propagating waves}
		Due to grating, the positive and the negative modes couple with each other via diffractive coupling
		\begin{equation}
			\label{eq:diffcoupl}
			\mathcal{H}^{\text{diffrac}}=\int \dd x\, U_D(x)\mathbf{\Phi}_+^\dagger(x)\mathbf{\Phi}_-(x)+ h.c.
		\end{equation}
		where the diffractive coupling function $U_D(x)$ is periodic with period $a$:
		\begin{equation}
			U_D(x)=\sum_{l\in \mathbb{Z}} U_l e^{i l K_0 x},
		\end{equation}
		where $U_l=U_{-l}$  because of the $C_2$ symmetry (reflection $x\to -x$) of the grating . Due to the diffractive coupling, effectively, the positive mode couple with the negative mode that is shifted by $lK_0$ in the momentum space. Vice versa, one can think of the diffractive coupling is the negative mode couples with the positive mode that is shifted by $-lK_0$ in the momentum space. The bandgaps will be open at the \textit{crossing points} between the positive (negative) band and the shifted negative (positive) band.  The strong coupling points are $K^+_C=\frac{lK_0}{2}$ of the positive band and $-K^-_C=-\frac{lK_0}{2}$ of the negative band. These are also the high symmetry points of the BZs. Due to the diffractive coupling mechanism, $l$ is called \textit{diffractive order}. We can rewrite the coupling in terms of the \textit{effective} wave-functions defined in Eqs. \eqref{eq:efwq},\eqref{eq:Fourier} and \eqref{eq:efwx}:
		
		\begin{equation}
			\label{eq:diff}
			\!\!\!\!	\mathcal{H}^{\text{diffrac}}=\!\!\!\!\!\!\sum_{\substack{l \in \mathbb{N}^*\\
					-l \leq n \leq l}}\!\!\!\!\!\! \int \dd x\, U_l \Phi_{l+n,+}^\dagger(x)\Phi_{l-n,-}(x)+ h.c.
		\end{equation}

		Note that since positive mode has positive wave-vectors and the negative mode has negative wave-vectors, only $l \in \mathbb{N}^*$ appears in the summation of Eq. \eqref{eq:diff} and $n$ runs from $-l$ to $l$ due to momentum conservation. However, the effective coupling becomes important when the energies of positive and negative bands are approximately identical, which corresponds to $n=0$. Hence we rewrite the diffractive Hamiltonian as:
		\begin{align}
			\label{eq:diffrac}
			\!\!\!\!\!	\mathcal{H}^{\text{diffrac}}=\sum_{l \in \mathbb{N}^*}\int \dd x\, U_l \Phi_{l,+}^\dagger(x)\Phi_{l,-}(x)+ h.c=\sum_{l \in \mathbb{N}^*}\int \frac{\dd  q}{2\pi}\, U_l \Phi_{l,+}^\dagger(q)\Phi_{l,-}(q)+ h.c
		\end{align} 
		Combining the \eqref{eq:Hfreeq} and the diffractive coupling \eqref{eq:diffrac}, we can derive the \textit{effective} Hamiltonian near the high symmetry point in the momentum basis
		\begin{small}
			\begin{equation}
				\label{eq:Hsingx}
				H_\text{single}(q)=		\mathcal{H}^{\text{free}}+\mathcal{H}^{\text{diffrac}}=\!\!\left( \begin{matrix}
					\omega_{0l}+v_l q& U_l  \\ U_l  & \omega_{0l}-v_l q
				\end{matrix} \right)
			\end{equation}	
		\end{small}
		From now on, we will concentrate on the high symmetry point corresponds to  $l=1$. We then replace $\omega_{01}\rightarrow \omega_0$, $U_1 \rightarrow U$ and $v_{1}\rightarrow v$, thus:
		\begin{equation}
			\label{eq:Hsingq}
			H_{\text{single}}(q)=\left( \begin{matrix}
				\omega_{0} + v q & U  \\ U  & \omega_{0} -  v q
			\end{matrix} \right)
		\end{equation}	
		In the subsequent sections ,we will obmit the $l$ indices and implicitly use $\Phi_{\pm}$ as $\Phi_{1,\pm}$ in \eqref{eq:Fourier}. 
		\subsection{Effective Hamiltonian of bilayer}
		\begin{figure}[hbt!]
			\begin{center}
				\includegraphics[width=0.35\textwidth]{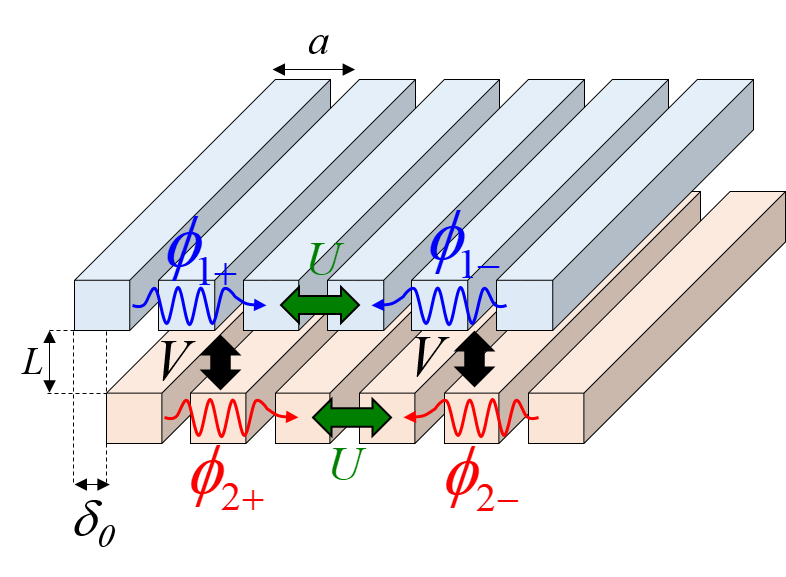}
			\end{center}
			\caption{(a) Sketch of a bilayer grating structure.}
			\label{fig:fishbone}
		\end{figure}
		
		To understand the inter-layer coupling mechanisms, an intuitive and informative example is the configuration of bilayer photonic lattice, referred to as the ``fish-bone'' structure in Ref~\cite{NHS2018}. Such a configuration consists of two identical gratings, one on top of the other with a relative displacement $\delta_0$ (Fig \ref{fig:fishbone}). Two special configurations of $\delta_0/a_0=0$ and 0.5 are respectively the equivalent of $AA$ and $AB$ stackings in Bilayer Graphene structure~\cite{Rozhkov2016}. We use the notation system  with the implementation of  index $(1)$ and $(2)$ to distinguish the \textit{upper} and \textit{lower} layer.
		We consider the basis made of \textit{effective} wave-functions near the crossing point of the positive and the negative bands of each layer
		\begin{equation}
			\label{eq:psi}
			\Psi^{(1)}(x)
			=\left(\begin{matrix}
				\Phi_{+}^{(1)}(x) \\ \Phi_{-}^{(1)}(x) 
			\end{matrix}\right),\quad
			\Psi^{(2)}(y)
			=\left(\begin{matrix}
				\Phi_{+}^{(2)}(y) \\ \Phi_{-}^{(2)}(y) 
			\end{matrix}\right).
		\end{equation}
		Similar to the case of single layer, the Hamiltonian densities of uncoupled layers in these basis are:
		\begin{small}
			\begin{equation}
				\label{eq:Hul}
				H_\text{single}^{(1)}(x)=\!\!\left( \begin{matrix}
					\omega_{0}-i v\d_x& U  \\ 
					U  & \omega_{0}+iv\d_x
				\end{matrix} \right),\quad
				H_\text{single}^{(2)}(y)=\!\!\left( \begin{matrix}
					\omega_{0}-iv\d_y& U  \\ 
					U  & \omega_{0}+iv\d_y
				\end{matrix} \right).
			\end{equation}
		\end{small}
		
		The evanescent coupling of the bilayer configuration is \footnote{Here, we ignore the coupling between the positive (negative) mode on the upper layer and the negative (positive) mode on the lower layer. This coupling includes a fast oscillation factor dues to the fact that the positive mode and the negative mode have different wave vectors.}
		\begin{equation}
			\!\!\!\!\!	\mathcal{H}^\text{inter}_{\text{bilayer}}=\int\dd x \int \dd y\,\left \{ \mathbf{\Phi}_{+}^{(1)\,\dagger}(x) \mathbf{\Phi}_{+}^{(2)}(y) \mathcal{V}_\text{f-b}(x-y)\right. \\+ \left. \mathbf{\Phi}_{-}^{(1)\,\dagger}(x) \mathbf{\Phi}_{-}^{(2)}(y) \mathcal{V}_\text{f-b}(x-y)\right \} + h.c
		\end{equation}
		In the regime in $L\ll a$, we can assume that $\mathcal{V}_\text{f-b}(x-y)=V.\delta(x-y-\delta_0)$.  If we only consider the effective modes near the symmetry point corresonds to $l=1$,  we rewrite the inter-layer coupling Hamiltonian as:
		
		\begin{equation}
			\label{eq:inter-layerH}
			\mathcal{H}^\text{inter}_{\text{bilayer}}=V\int\dd x\,\left\{\Phi_{+}^{(1) \dagger}(x)\Phi_{+}^{(2)}(x-\delta)e^{-i\frac{K_0}{2}\delta}+\Phi_{-}^{(1) \dagger}(x)\Phi_{-}^{(2)}(x-\delta)e^{i\frac{K_0}{2}\delta}\right\} + h.c 
		\end{equation}
		
		We then replace $\d_y \rightarrow \d_x$ and $y \rightarrow x-\delta_0$ in equation \eqref{eq:psi}  . The effective basis when working with both layers is given by: 
		\begin{equation}
			\label{eq:basisFBA}
			\Psi^{\text{bilayer}}(x)=\left(
			\begin{matrix}
				\Phi_{+}^{(1)}(x) \\ 
				\Phi_{-}^{(1)}(x) \\
				\quad\;\Phi_{+}^{(2)}(x-\delta_0) \\ 
				\quad\;\Phi_{-}^{(2)}(x-\delta_0) 
			\end{matrix}
			\right)
		\end{equation}
		for real space and momentum space respectively. The matrix representation of inter-layer coupling Hamiltonian of Eq.\eqref{eq:inter-layerH} in real space is written as:
		\begin{equation}
			\label{eq:Hinter}
			H^{\text{inter}}_{\text{bilayer}}(x)=\left(\begin{matrix}
				\bigzero_{2\times 2} & T_0\\
				T_0^{\dagger} & \bigzero_{2\times 2}
			\end{matrix} \right)\,,
		\end{equation}
		with the interlayer coupling matrix
		\begin{equation}
			\label{eq:Deltam}
			T_0=\left(\begin{matrix}
				Ve^{-i\frac{K_0}{2}\delta_0} & 0  \\
				0 & Ve^{i\frac{K_0}{2}\delta_0}
			\end{matrix} \right)\,.
		\end{equation}
		The bilayer Hamiltonian consists of the Hamiltonian of uncoupled layers and the inter-layer coupling Hamiltonian. Using effective Hamiltonians \eqref{eq:Hul} and the interlayer coupling \eqref{eq:Hinter}, we obtain the effective Hamiltonian for the bilayer system:
		\begin{equation}
			\label{eq:bilayer}
			H_{\text{bilayer}}=\left(\begin{matrix}
				H_\text{single} & T_0\\ 
				T_0^{\dagger} & H_\text{single} \end{matrix} \right),
		\end{equation}
		
		Another form of the bilayer Hamiltonian in momentum space 
		is reported in Ref.~\cite{NHS2018}. One can show that the two bilayer Hamiltonians are equivalent using a simple transformation of the basis.
		\subsection{Hamiltonian of the moire\'e bilayer}
		\begin{figure}[hbt!]
			\begin{center}
				\includegraphics[width=0.6\textwidth]{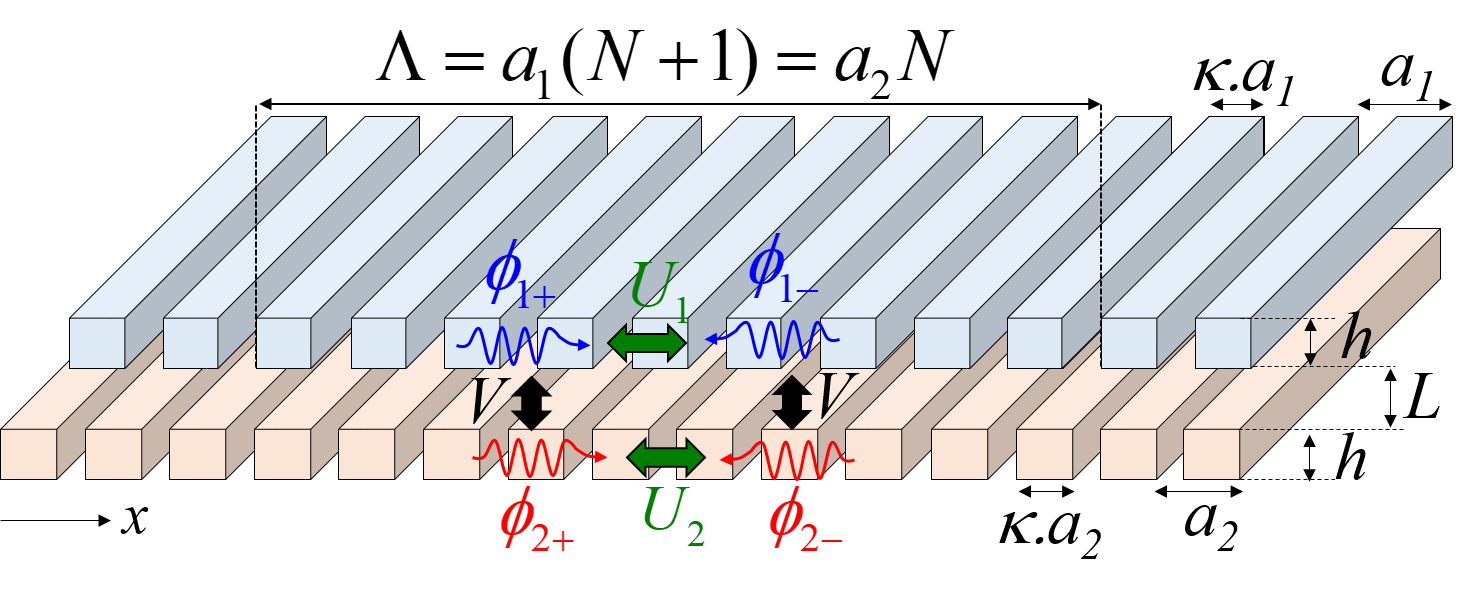}
			\end{center}
			\caption{(a) Sketch of a moir\'e structure.}
			\label{fig:moire}
		\end{figure}
		
		\subsubsection{Hamiltonian of uncoupled layers}
		
		We now consider a moir\'e bilayer of parameters as discussed in the main text (see Fig.~\ref{fig:moire}). With such geometrical design, the Hamiltonian of the uncoupled layers from moir\'e configuration has the same form as the ones of uncoupled layers as in the bilayer configuration. The only difference to the bilayer configuration is the mismatch between BZ-sizes of the two layers ($K_1=2\pi/a_1$ for the upper layer, and $K_2=2\pi/a_2$ for the lower layer). The decomposition of wavefunctions corresponding to positive and negative modes is given by:
		\begin{equation}
			\mathbf{\Phi}^{(1)}_{\pm}(x)
			=\sum_{l\in \mathbb{Z}}e^{\pm i \frac{lK_1}{2}x}\Phi^{(1)}_{l,\pm}(x)
			=\sum_{l \in \mathbb{Z}}e^{\pm i \frac{lK_1}{2}x}\int_{-\frac{K_1}{4}}^{\frac{K_1}{4}}\frac{\dd q}{2\pi}\, \Phi^{(1)}_{l,\pm}(q)e^{iqx},
			\label{eq:fullwave1}
		\end{equation}
		\begin{equation}
			\mathbf{\Phi}^{(2)}_{\pm}(y)
			=\sum_{l\in \mathbb{Z}}e^{\pm i \frac{lK_2}{2}y}\Phi^{(2)}_{l,\pm}(y)
			=\sum_{l \in \mathbb{Z}}e^{\pm i \frac{lK_2}{2}y}\int_{-\frac{K_2}{4}}^{\frac{K_2}{4}}\frac{\dd q}{2\pi}\, \Phi^{(2)}_{l,\pm}(q)e^{iqy}.
			\label{eq:fullwave2}
		\end{equation}
		Since the mismatch between BZ-sizes $K_M \ll K_1, K_2$, we expect the interlayer coupling to play an important role when the upper and lower modes are near the symmetry point with the same index $l$. We consider the effective theory near the symmetry point $l=1$, and omit the $l$ index by implicitly use $\Phi^{(1)}_{\pm}$ as $\Phi^{(1)}_{1,\pm}$ and $\Phi^{(2)}_{\pm}$ as $\Phi^{(2)}_{1,\pm}$. The basis made of \textit{effective} wave-functions near the crossing point of the positive and the negative bands of each layer
		\begin{equation}
			\!\!\!\!	\Psi^{(1)}(x)
			=\left(\begin{matrix}
				\Phi_{+}^{(1)}(x) \\ \Phi_{-}^{(1)}(x) 
			\end{matrix}\right),\quad
			\Psi^{(2)}(y)
			=\left(\begin{matrix}
				\Phi_{+}^{(2)}(y) \\ \Phi_{-}^{(2)}(y) 
			\end{matrix}\right).
		\end{equation}
		The Hamiltonian densities of uncoupled layers in these basis are:
		
		\begin{equation}
			\label{eq:Hsinxul}
			H_\text{single}^{(1)}(x)=\left( \begin{matrix}
				\omega_{0}^{(1)}-iv^{(1)}\d_x& U^{(1)}  \\ 
				U^{(1)}  & \omega_{0}^{(1)}+iv^{(1)}\d_x
			\end{matrix} \right)\, \quad
			H_\text{single}^{(2)}(y)=\left( \begin{matrix}
				\omega_{0}^{(2)}-i v^{(2)}\d_y & U^{(2)}  \\ 
				U^{(2)}  & \omega_{0}^{(2)}+iv^{(2)}\d_y  
			\end{matrix} \right)\,.
		\end{equation}
		
		The parameters of the Hamiltonians \eqref{eq:Hsinxul} are determined from the simulation and experiment fitting for a single-layer uni-dimensional photonic crystal slab.

		\subsubsection{Hamiltonian of inter-layer coupling: moir\'e configuration}
		\label{sec:derivationHmoire}
		
		Co-propagating waves of the same momentum but from different layers are coupled via evanescent coupling. The evanescent mechanism is written in term of the coupling Hamiltonian  as
		\begin{equation}
			\label{eq:HinterM}
			\mathcal{H}^\text{inter}=\int\dd x \int \dd y\,\left [ \mathbf{\Phi}_{+}^{(1)\,\dagger}(x) \mathbf{\Phi}_{+}^{(2)}(y) \mathcal{V}(x-y)+ \mathbf{\Phi}_{-}^{(1)\,\dagger}(x) \mathbf{\Phi}_{-}^{l}(y) \mathcal{V}(x-y)\right] + h.c
		\end{equation}
		When $L\ll a$, we can assume that $\mathcal{V}(x-y)=V\delta(x-y-\delta_0)$ where $\delta_0$ is the offset shift between the two layers. Moreover, as discussed in the main text, the value of $\delta_0$ is not relevant for the moir\'e structure, and we can assume it to be zero.  Considering the effective model near the symmetry points corresponding to $l=1$.  We rewrite the inter-coupling Hamiltonian \eqref{eq:HinterM} as the coupling of effective basis $\Phi^{(1)}_{\pm}(x)$ and $\Phi^{(2)}_{\pm}(x)$
		
		\begin{equation}
			\label{eq:inter-layerH_moire}
			\begin{split}
				\mathcal{H}^\text{inter}= &V\int\dd x\,\left[\Phi_{+}^{(1) \dagger}(x)\Phi_{+}^{(2)}(x)e^{-i\frac{(K_1-K_2)}{2}x}+\Phi_{-}^{(1) \dagger}(x)\Phi_{-}^{(2)}(x)e^{i\frac{(K_1-K_2)}{2}x}\right] + h.c \\
				=&V\int\dd x\,\left[\Phi_{+}^{(1) \dagger}(x)\Phi_{+}^{(2)}(x)e^{-i\frac{K_M}{2}x}+\Phi_{-}^{(1) \dagger}(x)\Phi_{-}^{(2)}(x)e^{i\frac{K_M}{2}x}\right] + h.c .
			\end{split}
		\end{equation}
		Some remarks are in order. We now understand the origin of the spatial dependent phase shift $\phi(x)$ in Eq. \eqref{eq:phi} in the main text by looking at the expansions \eqref{eq:fullwave1} and \eqref{eq:fullwave2}. Due to the mismatch between BZ-sizes, there is a different phase between the upper and the lower modes near the symmetry points corresponding to the same $m$. 
		We then choose an effective basis when working with both layers
		\begin{equation}
			\label{eq:basisM}
			\Psi^{\text{moir\'e}}(x)=\left(
			\begin{matrix}
				\Phi_{+}^{(1)}(x) \\ 
				\Phi_{-}^{(1)}(x) \\
				\quad\;\Phi_{+}^{(2)}(x) \\ 
				\quad\;\Phi_{-}^{(2)}(x) 
			\end{matrix}
			\right)
		\end{equation}
		
		We then replace $\d_y \rightarrow \d_x$ in Eq. \eqref{eq:Hsinxul} and obtain 
		the matrix representation of  the effective Hamiltonian of Eq.\eqref{eq:inter-layerH_moire} in the effective basis \eqref{eq:basisM}:
		\begin{equation}
			\label{eq:Hmoirex}
			H_{\text{moir\'e}}(x)=\left(\begin{matrix}
				H_\text{single}^{(1)}(x) & T(x)\\ 
				T^{\dagger}(x) & H_\text{single}^{(2)}(x)  
			\end{matrix} \right)\,.
		\end{equation}
		the interlay coupling matrix
		\begin{equation}
			\label{eq:Tm}
			T(x)=e^{-i\frac{K_M}{2}x}T_1 + e^{i\frac{K_M}{2}x}T_2, \quad
			T_1=\left(\begin{matrix}
				V  & 0 \\
				0& 0
			\end{matrix}\right), \qquad  T_2=\left(\begin{matrix}
				0 & 0 \\
				0& V 
			\end{matrix}\right)
		\end{equation}
		The difference of period would lead to a slight difference of values of the offset $\omega_0$ and the intra-layer coupling strength $U$ for each grating and a small modification of $V$ with respect to the case of Bilayer lattice. However, since $\omega_0 \gg U,V$, in the first approximation, only $\omega_0$ varies when switching from upper to lower layer.
		
		The decomposition \eqref{eq:Tm} shows two types of inter-layer coupling in momentum space: 
		\begin{itemize}
			\item The positive mode with effective momentum $q$ in the upper layer  will couple to the positive mode with effective momentum $q+\frac{K_M}{2}$ in the lower layer via $T_1$.
			\item The negative mode with effective momentum $q$ in the upper layer  will couple to the negative mode with effective momentum $q-\frac{K_M}{2}$ in the lower layer via $T_2$. 
		\end{itemize}
		We demonstrate this coupling mechanism in the momentum space explicitly in Fig \ref{fig:connect}b. This situation is similar to the inter-layer coupling model suggested by Bistrizer and Mac Donald in twisted bilayer graphene~\cite{Bistritzer:2010}; the only difference is that in twisted bilayer graphene, there are three couplings $T_1,T_2$ and $T_3$ corresponding to three momentum shifts instead of just two.\\
		
		\begin{figure}
			\begin{center}
				\includegraphics[width=1 \textwidth]{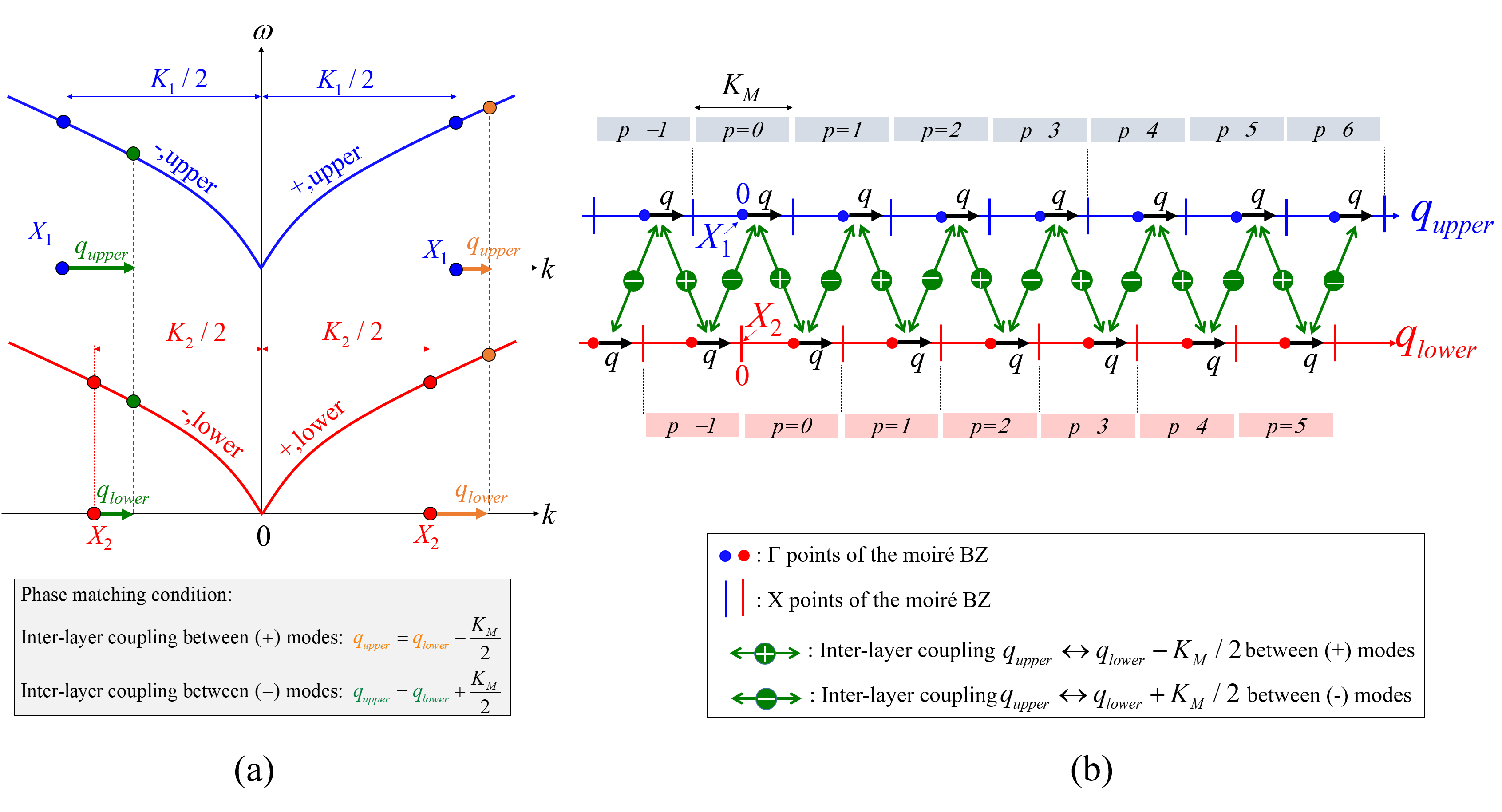}	
				\caption{(a)Phase-matching condition (conservation of momentum) for inter-layer coupling between co-propagating waves. (b) Inter-layer coupling mechanism in momentum space between different moir\'e B: Modes in the upper(lower) layer with Bloch momentum $q$ couple to modes in the lower(upper) layer with Bloch momentum $q-\frac{K_M}{2}$ and $q+\frac{K_M}{2}$. Each moir\'e BZ is indicated by its index $p$. }
				\label{fig:connect}
			\end{center}
		\end{figure}
		
		{\it\textbf{A change of basis:}} 
		
		The effective basis \eqref{eq:basisM} was chosen in the same manner as in the twisted bilayer graphene literature~\cite{Tarnopolsky2019}. Consequently, the Hamiltonian \eqref{eq:Hmoirex} shares the same pattern as the Hamiltonian derived by  Bistritzer and MacDonald in Ref \cite{Bistritzer:2010,Tarnopolsky2019} as expected. Notice that in the effective basis \eqref{eq:basisM}, the origins of the effective momenta are different. The wave-function in the coordinate space is given by  
		
		\begin{equation}
			\label{eq:basisReal}
			\mathbf{\Psi}(x)=\left(
			\begin{matrix}
				e^{i\frac{K_1}{2}x}\Phi_{+}^{(1)}(x) \\ 
				e^{-i\frac{K_1}{2}x} \Phi_{-}^{(1)}(x) \\
				\quad\;e^{i\frac{K_2}{2}x}\Phi_{+}^{(2)}(x) \\ 
				\quad\;e^{-i\frac{K_2}{2}x}\Phi_{-}^{(2)}(x) 
			\end{matrix}
			\right).
		\end{equation}
		The wave-function of the electromagnetic wave near the vicinity of $X$  point on the upper layer and lower layer can be read off from \eqref{eq:basisReal} as 
		\begin{equation}
			\label{eq:wf}
			\psi^{up}(x)=e^{i\frac{K_1}{2}x}\Phi_{+}^{(1)}(x)+e^{-i\frac{K_1}{2}x}\Phi_{-}^{(1)}(x),\quad \psi^{low}(x)=e^{i\frac{K_2}{2}x}\Phi_{+}^{(2)}(x)+e^{-i\frac{K_2}{2}x}\Phi_{-}^{(2)}(x).
		\end{equation}
		One can use \eqref{eq:wf} solved from effective Hamiltonian \eqref{eq:Hmoirex} to compare directly with the electromagnetic wave in coordinate space of simulations and experiments. However, it is helpful to introduce another effective basis such that the wavefunction in the coordinate space is 
		\begin{equation}
			\label{eq:basisReal1}
			\mathbf{\Psi}(x)=\left(
			\begin{matrix}
				e^{i\frac{K_1}{2}x}\Phi_{+}^{(1)}(x) \\ 
				e^{-i\frac{K_1}{2}x} \Phi_{-}^{(1)}(x) \\
				\quad\;e^{i\frac{K_2}{2}x}\Phi_{+}^{(2)}(x) \\ 
				\quad\;e^{-i\frac{K_2}{2}x}\Phi_{-}^{(2)}(x) 
			\end{matrix}
			\right)=\left(
			\begin{matrix}
				e^{i\frac{K_1}{2}x}\Phi_{+}^{(1)}(x) \\ 
				e^{-i\frac{K_1}{2}x} \Phi_{-}^{(1)}(x) \\
				\quad\;e^{i\frac{K_1}{2}x}\tilde{\Phi}_{+}^{(2)}(x) \\ 
				\quad\;e^{-i\frac{K_1}{2}x}\tilde{\Phi}_{-}^{(2)}(x) 
			\end{matrix}
			\right),
		\end{equation}
		which implies 
		\begin{equation}
			\tilde{\Phi}_{+}^{(2)}(x)=e^{-iK_M x/2} \Phi_{+}^{(2)}(x), \quad \tilde{\Phi}_{-}^{(2)}(x)=e^{iK_M x/2} \Phi_{-}^{(2)}(x).
		\end{equation}
		We are able to rewrite the moir\'e Hamiltonian \eqref{eq:Hmoirex} in the new effective basis 
		\begin{equation}
			\label{eq:basisM1}
			\tilde{\Psi}^{\text{moir\'e}}(x)=\left(
			\begin{matrix}
				\Phi_{+}^{(1)}(x) \\ 
				\Phi_{-}^{(1)}(x) \\
				\quad\;\tilde{\Phi}_{+}^{(2)}(x) \\ 
				\quad\;\tilde{\Phi}_{-}^{(2)}(x) 
			\end{matrix}
			\right)
		\end{equation}
		explicitly as follow
		\begin{align}
			\label{eq:Hmoirex1}
			H &= \left(\begin{matrix}
				-v^{(1)}i\partial_x + \omega_0^{(1)}  & U^{(1)} & V  & 0 \\
				U^{(1)} & +v^{(1)}i\partial_x + \omega_0^{(1)} & 0 & V  \\
				V  & 0 & -v^{(2)}i\partial_x + \omega_0^{(2)} & U^{(2)}e^{-i K_M x} \\
				0 & V  & U^{(2)}e^{i K_M x} & +v^{(2)}i\partial_x + \omega_0^{(2)}
			\end{matrix}\right).
		\end{align}
		Since $K_1=(N+1)K_M$, the momentum of the effective basis $\Phi^{(1)}_\pm(q)$ and $\tilde{\Phi}^{(2)}_\pm(q)$ are folded back to the same point in the moir\'e BZ. Therefore, the new effective basis \eqref{eq:basisM1} is convenient to compare with the moir\'e wave-functions from simulations and experiments in the momentum space (moir\'e BZ). The Hamiltonian \eqref{eq:Hmoirex1} is nothing but the effective Hamiltonian (1) in the main text. 
		
		\subsection{Qualitative analysis of the effective Hamiltonian}
		\label{sec:qualitative}
		\subsubsection{Dimensional analysis and simplified model}\label{sec:dimension}
		
		Let us notice that when a time scale (or equivalently energy, or frequency, scale) is fixed, one is still free to choose a length scale in the Hamiltonian~\eqref{eq:Hmoirex}. To fix a length scale, one can set $v=1$. Since one can choose an arbitrary reference value for the energy, clearly the absolute values of $\omega^{(1)}_{0}$ and $\omega^{(2)}_{0}$ are not important. It is however crucial that they are different to separate the energy bands of the two uncoupled layers from each other. We thus can substitute $\omega^{(1)}_{0} \to \Delta$, $\omega^{(2)}_{0} \to -\Delta$ for the qualitative consideration, i.e.,  chosing the zero-energy to be $\omega_{00}=\left(\omega^{(1)}_{0}+\omega^{(2)}_{0}\right)/2$. Furthermore, let $U = (U^{(1)} +U^{(2)})/2$ and $\Delta_U = (U^{(1)} - U^{(2)})/2$. 
		We then have the simplified Hamiltonian as
		\begin{align}
			\begin{split}\label{eq:simplified_MoireH}
				H =&-i \partial_x  (\II \otimes \sigma_z) + \Delta (\sigma_z \otimes \II) + U (\II \otimes \sigma_x) \\
				&+ V \left(\sigma_+ \otimes e^{-i q_0 x\sigma_z} + \sigma_- \otimes e^{+i q_0 x \sigma_z}\right) \\
				&+ \Delta_U \sigma_z \otimes \sigma_x, 
			\end{split}
		\end{align}
		where the characteristic wavevector is $q_0=K_M/2$ with $K_M$ is the moir\'e wavevector. Here $\otimes$ denotes the Kronecker product, $\sigma_{x,y,z}$ are Pauli matrices defined by $\sigma_x= \left(\begin{matrix} 0 &1 \\ 1 &0 \end{matrix}\right), \quad \sigma_y= \left(\begin{matrix} 0 &-i \\ i &0 \end{matrix}\right), \quad \sigma_z= \left(\begin{matrix} 1 &0 \\ 0 &-1 \end{matrix}\right)$, and $\sigma_\pm=(\sigma_x\pm i\sigma_y)/2$. The last term in Hamiltonian~\eqref{eq:simplified_MoireH} only leads to minor quantitative corrections; for qualitative analysis, one can set $\Delta_U=0$. We see then that the equation~\eqref{eq:simplified_MoireH} is characterised  by parameters $(U, \Delta, q_0, V)$. All of these quantities have the same dimension of energy (since $v=1$). One can effectively set one of them, e.g., $U$, to be the unit.
		
		Moreover, when we specialise to the particular realisation of the effective Hamiltonian~\eqref{eq:simplified_MoireH}, as in Appendix~\ref{sec:parameter_retrieval_single}, we see that the parameters $\Delta=\omega^{(1)}-\omega^{(2)}$, $\Delta_U=U^{(1)}-U^{(2)}$  and $q_0=\frac{K_M}{2}$ are in fact physically dependent through the straining parameter in the system. In this case, we therefore only have three independent physical parameters $(U, q_0, V)$. The model is specified by two dimensionless ratios between the independent parameters.


		\subsubsection{Periodicity and the Bloch Hamiltonian}
		\label{sec:effectiveHperiod}

		It is perhaps surprising when one notices that the Hamiltonian~\eqref{eq:simplified_MoireH} seem to be periodic with the double supercell period $2 \pi/ q_0=2 \Lambda$, which we refer to as \emph{apparent period}. Accordingly, naively solving these Hamiltonian one obtains a band structure with the \emph{apparent Brillouin zone} of size $K_M/2$. The Hamiltonian is in fact of higher translational symmetry. Indeed, let $T_{\Lambda}$ be the translation operator of one moir\'e period. Then one can easily verify that~\footnote{This can be demonstrated by using $\sigma_z \sigma_{\pm} \sigma_z^{\dagger} = -\sigma_{\pm}$ and $T_{\Lambda} e^{-i \theta \sigma_z} T_{\Lambda}^{\dagger} = -e^{-i \theta \sigma_z}$.} the Hamiltonian is invariant under the generalized translational operator $T_{\Lambda} (\sigma_z \otimes \II)$. 
		The operator $T_{\Lambda} (\sigma_z \otimes \II)$ generates the commutative group of generalised translational operators, under which the Hamiltonian is invariant. This shows that the \emph{actual period} of the system is, not surprisingly, the moir\'e period $\Lambda$.
		
		
		With the apparent period of $2 \Lambda$, the Bloch theorem states that we can assume the eigenstate of the Hamiltonian~\eqref{eq:simplified_MoireH} to be of the form 
		\begin{equation}
			\Psi(x)=e^{i q x} u_q(x),
			\label{eq:blochthr}
		\end{equation}
		where $q$ is the moir\'e Bloch vector, $ - q_0/2 \le q \le  +q_0/2$ and the four-spinor $u_q(x)$ is periodic with the apparent period $2\Lambda$.
		This leads to the Bloch Hamiltonian for $u_q(x)$,
		
		\begin{equation}
			\label{eq:HMoire_Bloch}
			H_q = 
			\begin{pmatrix}
				-i \partial_x + q + \Delta  & U^{(1)} & V e^{-i q_0 x} & 0 \\
				U^{(1)} & +i \partial_x - q + \Delta & 0 & V e^{+i q_0 x} \\
				V e^{+i q_0 x} & 0 & -i  \partial_x + q - \Delta & U^{(2)} \\
				0 & V e^{-i  q_0 x} & U^{(2)} & +i \partial_x - q - \Delta
			\end{pmatrix}.
		\end{equation}
		
		This Hamiltonian is to be solved for eigenvalues $E_q$  with periodic eigenstates $u_{q}(x)$, where the latter is also denoted by $u_{E,q}(x)$ when the explicit energy value is necessary for the clarity. The periodicity of the Bloch wavefunction $u_q(x)$ allows for the solution of the eigenvalue problem to be found through Fourier expansion.
		
		It is important to emphasize again that when using the apparent period $2 \Lambda$ of the Hamiltonian to calculate the band structure, the Bloch momentum $q$ in equation~\eqref{eq:HMoire_Bloch} is folded within $[-q_0/2,q_0/2]$. In order to unfold the band to the full moir\'e Brillouin zone $[-q_0,q_0]$, one simply solves the Bloch Hamiltonian for $q$ in the full moir\'e Brillouin zone, but maintains only solutions that satisfy the generalised Bloch theorem $T_{\Lambda} (\sigma_z \otimes \II) \Psi (x) =  e^{iq x} \Psi (x) $. In this way, the unfolded band structure such as in Fig.~\ref{fig:band} can be obtained.
		
		
		
		\subsubsection{Symmetry analysis}
		\label{sec:effectiveHsymmetry}
		Since the moir\'e system has spatial refection and time-reversal symmetries, one expects that the Hamiltonian~\eqref{eq:simplified_MoireH} also carries these symmetries. This is indeed the case:
		\begin{itemize}
			\item 	{\it\textbf{Spatial reflection:}} Consider the reflection along the $x$-axis. Let $P$ denote the pure spatial coordinate reflection operator.  Since the reflection of the $x$-axis also changes the signs of the momenta within each chain, it also exchanges the two basis wavefunctions chosen is Section~\ref{sec:derivationHmoire}. Therefore one can expect that the full reflection operator to be $(\II \otimes \sigma_x) P$. One can easily verify that the Hamiltonian~\eqref{eq:simplified_MoireH} is indeed invariant under this full reflection operator $(\II \otimes \sigma_x) P$. As also expected, the spatial reflection $(\II \otimes \sigma_x) P$ brings the Bloch Hamiltonian $H_{q}$ in Eq.~\eqref{eq:HMoire_Bloch}, to $H_{-q}$, implying the energy bands are symmetric under reflecting the Bloch wavevector, $E_q = E_{-q}$, and the Bloch wave functions obey $u_{E,q} (x) = u_{E,-q} (-x)$.
			
			\item {\it\textbf{Time reversal:}} Let $K$ be the complex conjugation. One can verify that the Hamiltonian~\eqref{eq:simplified_MoireH} is invariant under the full time reversal operator $(\mathbb{I} \otimes \sigma_x) K$. Again the time reversal operator brings $H_{q}$ to $H_{-q}$, implying the energy bands are symmetric under reflecting the wavevector, $E_q = E_{-q}$, and the Bloch wave functions obey $u_{E,q} (x) = u_{E,-q}^{\ast} (x)$.
		\end{itemize}


		

		
		\section{Some more details of the analysis of the effective Hamiltonian: flatbands, localization, tunnelling and bound states}
		
		\label{sec:effectiveHanalysis}
		\subsection{Fourier transform of the band structure}

		Being even and periodic with respect to the moir\'e wavevector $K_M=2 q_0$, an energy band $E(q)$ is completely described by Fourier coefficients $f_p = \frac{1}{2 q_0} \int_{-q_0}^{+q_0} d q \cos (p \pi/q_0) E(q)$. Fig.~\ref{fig:bandstructure} presents the Fourier components $f_p$ of the flat band as indicated in Fig.~\ref{fig:band} in the main text, but with slightly different dimensionless parameters as indicated in the caption. It shows that the first coefficients of the Fourier dominate over higher Fourier components, suggesting in an effective tight-binding model, the nearest neighbour coupling dominates. Importantly, higher Fourier coefficients, although small, do not vanish when the first coefficient vanishes (near the flat band). This suggests that the band, although becomes highly flat at the magic coupling, is not \emph{perfectly} flat.
		
		\begin{figure}[hbt!]
			\begin{center}
				\includegraphics[width=0.5\textwidth]{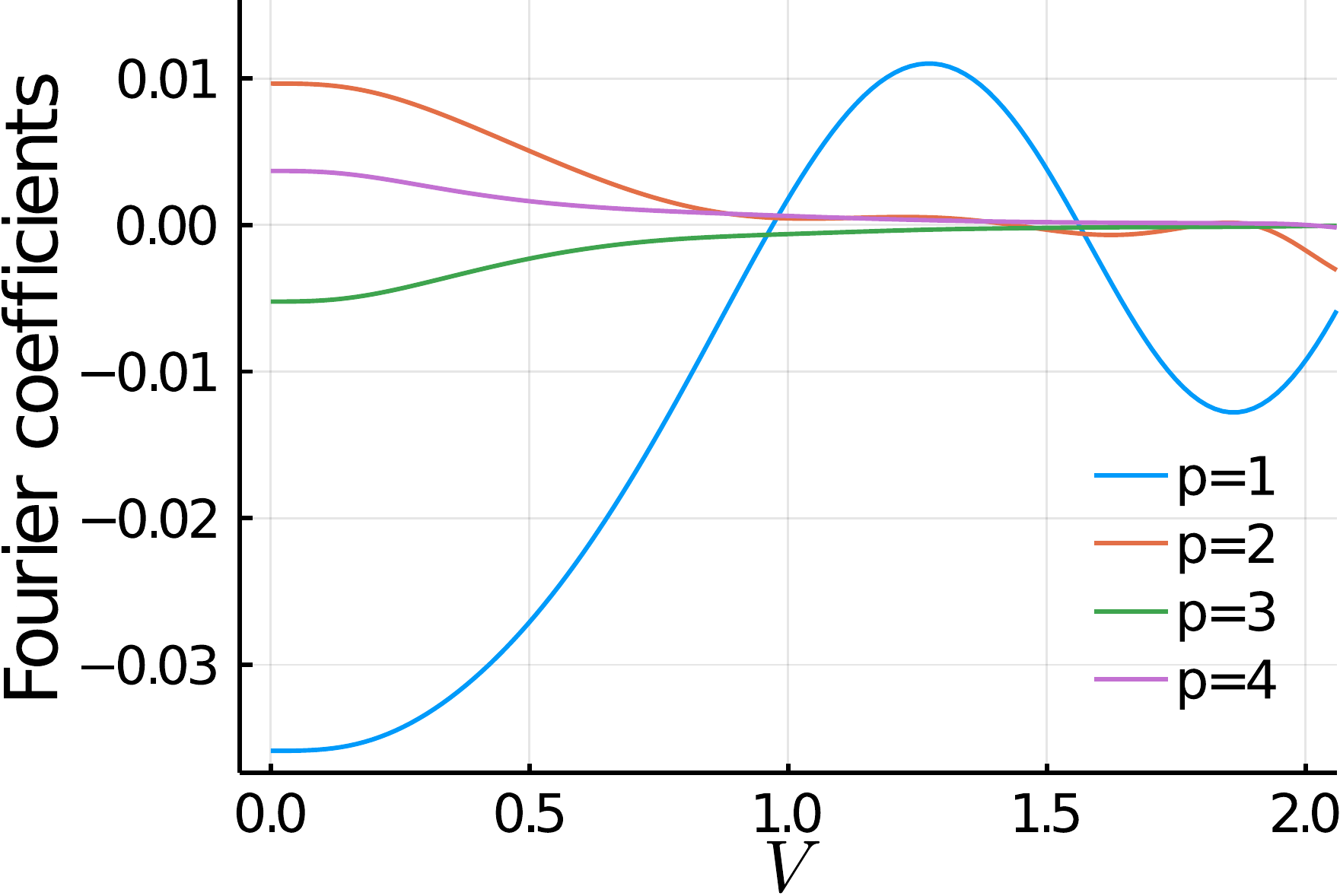}
			\end{center}
			\caption{The Fourier coefficients $f_p= \frac{1}{2q_0} \int_{-q_0}^{+q_0} d q \cos (p \pi/q_0) E(q)$ as functions of the couplings $V$ with $U=1$, $\Delta=-0.3$, $q_0=0.6$.}
			\label{fig:bandstructure}
		\end{figure}
		
		\subsection{Probability density distribution of the Bloch wave functions and Wannier functions}\label{sec:WannierCalculation_appendix}
		To study the Bloch wave function near the flat transition, we compute the density
		\begin{equation}
			\rho(x) = u_q(x)^{\dagger} u_q (x). 
		\end{equation}
		Figure~\ref{fig:wanniers} (left) demonstrates the probability densities of the Bloch wavefunctions with varying coupling $V$ across a flat transition. It is important to notice that while the Bloch wave functions tend to concentrate within a moir\'e period, they do not vanish anywhere (also when the band is flat). In particular, there is no qualitative change in the density of the Bloch wave function as the band is crossing the flat transition. 
		
		To consider the possibility of concentrating light in the moir\'e lattice, we compute the Wannier functions for the Bloch Hamiltonian~\eqref{eq:HMoire_Bloch}.
		The computation of the Wannier function requires fixing the arbitrary phase in the numerical solution of the eigenvectors of the Bloch Hamiltonian~\eqref{eq:HMoire_Bloch}.
		This is a known difficulty in computing Wannier functions with maximal localization~\cite{Vanderbilt2018a}. Fortunately, in one-dimensional systems, there is a known gauge fixing procedure, the \emph{twisted parallel transport gauge}, that allows for the computation of Wannier functions of maximal localization~\cite{Vanderbilt2018a}. 
		
		Upon fixing the twisted parallel transport gauge, the Wannier function is then obtained directly as 
		\begin{equation}
			W_0 (x) = \frac{1}{2 q_0} \int_{-q_0}^{+q_0} \mathrm{d} q e^{i qx} u_q(x).
			\label{eq:wannier}
		\end{equation}
		Notice that the integral runs over the full moir\'e BZ $[-q_0,q_0]$, that is, twice as much of the apparent BZ $[-q_0/2,q_0/2]$.
		Figure~\ref{fig:wanniers} (right) plots the probability density of the Wannier wavefunction~\eqref{eq:wannier}. While being highly concentrated, see Fig~\ref{fig:wanniers_confinement}, one should notice that the Wannier function extends beyond a single Moir\'e period. There is also no qualitative change in the density of the Wannier wave function as the band is crossing the flat transition. 

		\subsection{Dynamical signature of flatbands}
		
		From the above analysis, it is clear that the concentration of the probability density of the Bloch wave function or the Wannier function is not the signature of the flat band. In fact, the very physical meaning of localization in this context is a dynamic one.

		Suppose the system has a flatband $u_q(x)$, that is for some energy level $E_q=E_0$, independent of $q$. Then this is nothing but saying that $\Psi_q(x)= e^{i qx} u_q(x)$ are having the same energy $E_q=E_0$ for all $q$. This means that given any wave function in momentum space $v_q$, the wave packet
		\begin{equation}
			\Psi(x) = \sum_{q} v_q e^{i qx} u_q(x) 
		\end{equation}
		is also an eigen-wavefunction with energy $E_0$. As a consequence, the probability density $\Psi(x)^{\dagger} \Psi(x)$ is unchanged overtime.  This is true for any wavepacket $v_q$ in the Bloch momentum space, in particular the Wannier function~\eqref{eq:wannier}. 

		\begin{figure}
			\begin{center}
				\includegraphics[width=0.5\textwidth]{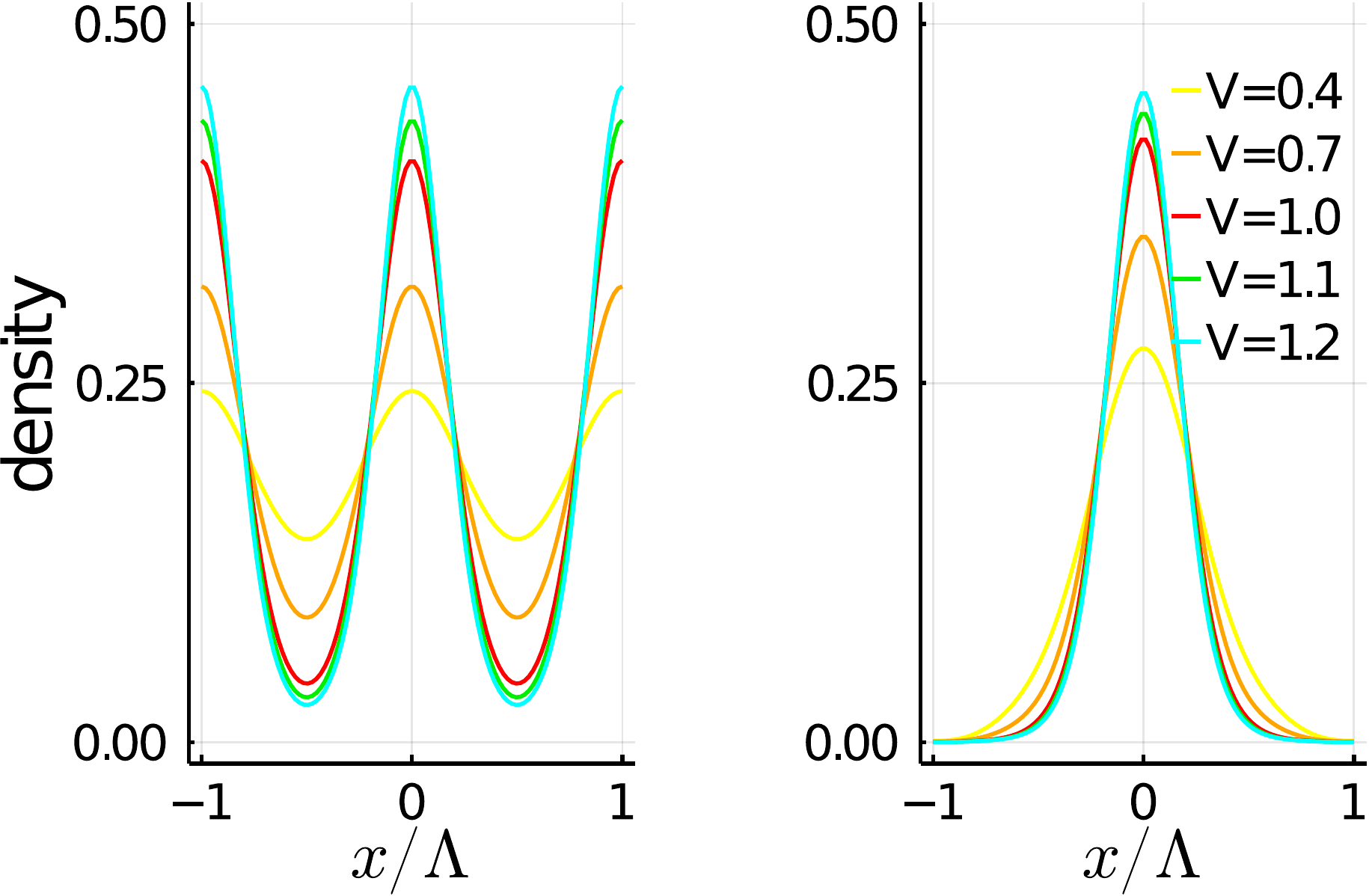}
			\end{center}
			\caption{Probability density of the Bloch wavefunction at  wavevector $q=0$ (left) and the Wannier functions (right) of the first positive band with parameters $U=1$, $\Delta=-0.3$, $q_0=0.6$, and varying coupling $V=0.1,0.7,0.9,1.1,1.3$ (with $V=1$ (red) near a flatband transition).}
			\label{fig:wanniers}
		\end{figure}
		
		\begin{figure}
			\begin{center}
				\includegraphics[width=0.5\textwidth]{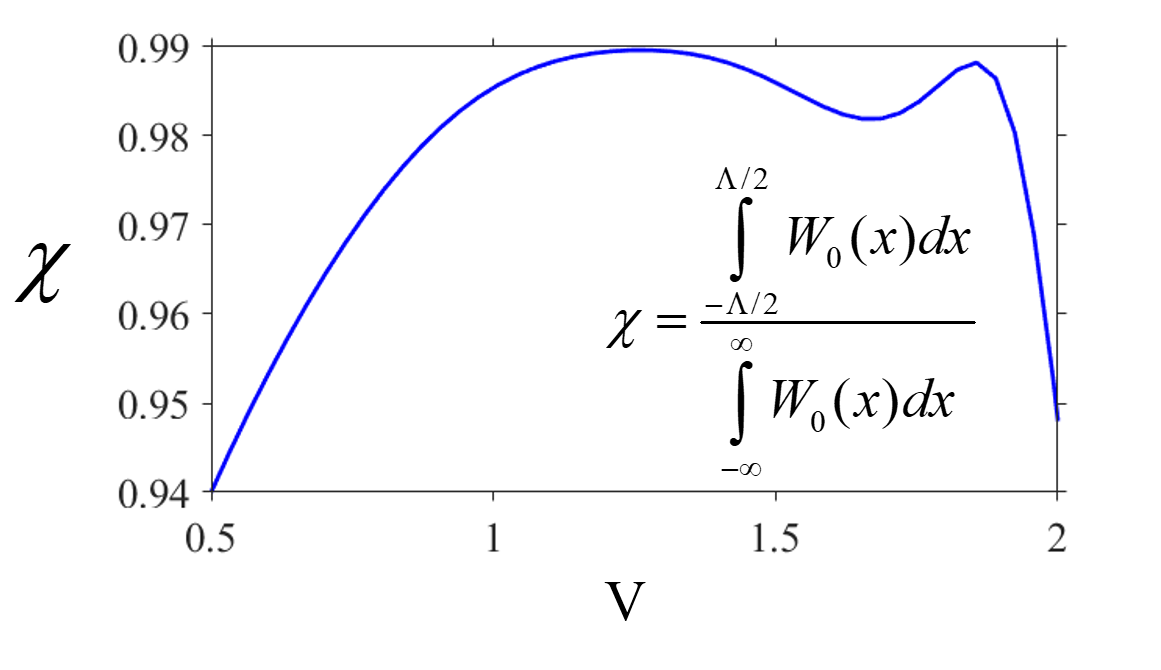}
			\end{center}
			\caption{The fraction of Wannier function confined within a moir\'e period.}
			\label{fig:wanniers_confinement}
		\end{figure}
		
		\subsection{Finite systems: tunnelling and resonances, bound states}\label{sec:Finite_system_appendix}
		To understand better the nature of the flat bands, we compute the tunnelling and bound states of light in a \emph{finite} number of moir\'e periods. This calculation can be carried out employing the (generalised) transfer matrix method~\cite{Davies1998a}, particularly adapted to the case of Dirac-like equations in Ref.~\cite{Nguyen2008a}.
		
		To do so, we rewrite the eigvenvalue equation
		\begin{equation}
			H \Psi (x) = E \Psi (x).
		\end{equation}
		into the form
		\begin{equation}
			\partial_x \Psi (x) = \mathcal{H} (x) \Psi(x)
			\label{eq:x-dynamics}
		\end{equation}
		where $\mathcal{H} (x)$ is a $4 \times 4$ matrix given by
		\begin{equation}
			\mathcal{H} (x) =
			i
			\begin{pmatrix}
				-(\Delta-E) & -U & -V e^{-i q_0 x} & 0 \\
				U & \Delta-E & 0 & V e^{+i q_0 x} \\
				-V e^{+i q_0 x} & 0 & -(\Delta-E) & -U \\
				0 & V e^{-i q_0 x} & U & \Delta-E
			\end{pmatrix}.
		\end{equation}
		All possible $x$-evolutions of the $x$-dynamical equation~\eqref{eq:x-dynamics} is described by the $4 \times 4$ $x$-evolution  operator $G_E(x_2,x_1)$, which is the solution of
		\begin{equation}
			\partial_{x_2} G_E (x_2,x_1) = \mathcal{H} (x_2) G_E (x_2,x_1),
		\end{equation}
		subject to the initial condition $G_E (x_2,x_1)= \mathbb{I}$.
		
		The function $G_E(x_2,x_1)$ summarises \emph{all} information about the eigenwave function $\Psi(x)$ corresponding to the eigenvalue $E$ of the Hamiltonian $H$. Therefore it is a convenient way to relate different properties of $H$, such as the existence of extended states, transmission amplitudes, probability distribution, the density of states, etc. On the other hand, with well-developed methods for the ordinary differential equations (ODEs)~\cite{Rackauckas2017a}, the computation of $G_E(x_2,x_1)$ is relatively easy. One should, however, notice that the $x$-dynamics is non-hermitian and sometimes numerical instabilities have to be addressed.
		
		\subsubsection{Boundary condition and the computation of tunnelling rate}
		To investigate the tunnelling phenomena through the finite moir\'e structure between $x_1$ and $x_2$, one has to consider the realisation of the asymptic area outside the moir\'e structure. For convenience, we choose this to be of the type of fishbone structure~\cite{NHS2018}; that is, fixing the phase in the coupling between the two chains in the Hamiltonian~\eqref{eq:simplified_MoireH} to be $e^{\pm iq_0x_1}$ (constant) for $x \le x_1$, and $e^{\pm iq_0x_2}$ (constant) for $x \ge x_2$.
		
		For the fixed phases $e^{\pm iq_0x_1}$ or $e^{\pm iq_0x_2}$, the eigenstate of the Halmitonian~\eqref{eq:simplified_MoireH} can be easily solved, resulted in the fishbone band structure~\cite{NHS2018}. Plugging a plane-wave solution $C e^{ik x}$ into the resulted Hamiltonian, one finds the fishbone eigenvalue equation,
		\begin{equation}
			\begin{pmatrix}
				k + \Delta - E & U & Ve^{-i \phi} & 0 \\
				U & -k + \Delta  - E & 0 & V e^{+i \phi} \\
				V e^{+i \phi} & 0 & k - \Delta - E& U \\
				0 & V e^{-i \phi} & U & -k  - \Delta - E
			\end{pmatrix} C = 0,
			\label{eq:eigen_wavevector}
		\end{equation} 
		where  $\phi = q_0 x_1$ or $\phi = q_0 x_2$, which are here simply constants. Fixing the energy $E$, we are interested in solving this equation for $k$. The resulted equation is a generalised eigenvalue problem. In general, the obtained generalised eigenvalues $k$ are complex. To fix an ordering, we order the four (generalised) eigenvalues $k$ according to their increasing phases, that is, the angles with respect to the real axis, computed counterclockwise.  
		
		Let us consider the possible solutions of equation~\eqref{eq:eigen_wavevector}. One sees that if $k$ is a solution, $k^{\ast}$ is also a solution (time-reversal symmetry). Also, if $k$ is a solution, $-k$ is also a solution (spatial reflection symmetry). In general, one has $4$ different wavevectors satisfying~\eqref{eq:eigen_wavevector}. If one of the solution $k$ is generically complex (i.e., not pure real or pure imaginary), then by acting with the time-reversal symmetry and reflection symmetry, one obtains all the other three solutions $k^{\ast}$, $-k$, $-k^{\ast}$, which are also generically complex. On the other hand, if one of the solution $k$ is real, then the time-reversal symmetry and the reflection symmetry only give $-k$ as another solution. There are then two possibilities: the other two solutions can also be real, or they must be purely imaginary. 
		
		To consider the tunelling phenomena, we are interested in the energy range of $+\Delta + U \le E \le -\Delta + U$ (for $\Delta < 0$). Here for a fixed energy $E$, there are two real wavevectors $ \pm k$ (with the convention $k \ge 0$), corresponding to the phases of $0$ and $\pi$. 
		Two other modes are of pure imaginary wavevectors $i\kappa$  (with the convention $\kappa \ge 0$) corresponding to exponential decaying or exponential amplifying modes and phases of $\pi/2$ and $3 \pi/2$.
		By $W$ we denote the matrix of which the columns are the corresponding eigenvectors (ordered such that phases of the eigenvalues increase, here must be $0$,$\pi/2$,$\pi$ and $3 \pi/2$). The general wavefunction depends on $4$ amplitudes of these different solutions, $a^\pm$ and $b^{\pm}$, explicitly given by
		\begin{equation}
			\Psi(x) = W V (x) 
			\begin{pmatrix}
				a^+ \\ 
				b^+ \\ 
				a^- \\
				b^-
			\end{pmatrix}
		\end{equation}
		where 
		\begin{equation}
			V(x) = 
			\begin{pmatrix} 
				e^{ikx} & 0 & 0 & 0 \\
				0 & e^{-\kappa x} & 0 & 0 \\
				0 & 0 & e^{-ik x} & 0 \\ 
				0 & 0 & 0 & e^{+\kappa x} 
			\end{pmatrix}.
		\end{equation}
		According to the ordering convention,  $a^{\pm}$ are the amplitudes of the travelling modes (corresponding to phases of eigenvalues $k$ of $0$ and $\pi$) and $b^{\pm}$ are the amplitudes of the exponential modes (corresponding to phases of the eigenvalues $k$ of $\pi/2$ and $3 \pi/2$). 
		
		This solution can be applied to both the areas $x \le x_1$ and $x \ge x_2$ with corresponding amplitudes $a^{\pm}_{1}$ and  $b^{\pm}_{1}$ and $a^{\pm}_{2}$ and  $b^{\pm}_{2}$. This results in the wave function at $x=x_1$ to be $\Psi (x_1) = W_1 V_1 (x_1) (a^+_1,b^+_1,a^-_1,b^-_1)^T$ and  at $x=x_2$ to be $\Psi (x_2) = W_2 V_2 (x_2) (a^+_2,b^+_2,a^-_2,b^-_2)^T$. Now using the solution of the wavefunction $G(x_2,x_1)$ through the moir\'e periods as obtained from the generalised transfer matrix, $\Psi(x_2) = G(x_2,x_1) \Psi(x_1)$, one obtains 
		\begin{equation}
			\mathcal{T}
			\begin{pmatrix}
				a^+_1 \\  
				b^+_1 \\
				a^-_1 \\
				b^-_1
			\end{pmatrix}
			=
			\begin{pmatrix}
				a^+_2 \\  
				b^+_2 \\
				a^-_2 \\
				b^-_2
			\end{pmatrix}
		\end{equation}
		where the \emph{transfer matrix} $\mathcal{T}$ is given by
		\begin{equation}
			\mathcal{T} = V_2(x_2)^{-1} W_2^{-1} G(x_2,x_1) W_1 V_1(x_1).
		\end{equation}
		To obtain the tunnelling rate, we apply the boundary condition $(a^+_1,b^+_1,a^-_1,b^-_1) = (1,0,r,l_1)$ and $(a^+_2,b^+_2,a^-_2,b^-_2) = (t,l_2,0,0)$. It is interesting to notice that the exponential modes also participate in the process: by injecting a plane wave at $a^{+}_1=1$, a wave is reflected at $a^-_1=r$, some part $a^+_2=t$ is transmitted though; and at the same time the (left and right) exponentially decaying modes are excited with amplitudes $l_1$ and $l_2$. This gives rise to the formula for the reflection coefficients and transmission coefficients as
		\begin{align}
			r &= - \frac{\mathcal{T}_{44} \mathcal{T}_{31} - \mathcal{T}_{34} \mathcal{T}_{41}}{\mathcal{T}_{33} \mathcal{T}_{44} - \mathcal{T}_{34} \mathcal{T}_{43}}  \\
			l_1 &=  -\frac{-\mathcal{T}_{43} \mathcal{T}_{31} + \mathcal{T}_{33} \mathcal{T}_{41}}{\mathcal{T}_{33} \mathcal{T}_{44} - \mathcal{T}_{34} \mathcal{T}_{43}}  \\
			t & = \mathcal{T}_{11} + r \mathcal{T}_{12} + l_1 \mathcal{T}_{14}.
		\end{align}
		Obtaining the transmission coefficients, one can exact its resonant structure, which indicates the quasi-bound states of light in the system. These obtained quasi-bound states can be compared to the band structure of the system of an infinite number of periods. However, it is even more convenient to study the exact bound states in a system of a finite number of moir\'e periods for our consideration.
		
		
		\subsubsection{Boundary condition and the computation of bound states}
		As for bound states, we consider again the asymptotic areas to be of fishbone type, but now at $x_1=\Lambda/2$ and $x_2=\Lambda/2+ p \Lambda$ for a integer number $p$. In this scenario, in the energy interval $U + \Delta \le E \le U - \Delta$ (notice again that $\Delta <0$), there is no extended states in the fishbone areas; all four wavevectors as solutions of ~\eqref{eq:eigen_wavevector} are generically complex. 
		Recall that we order the eigenvalues according to their angles with the real axis.
		To have a bound state, we apply the boundary condition for the amplitudes $(0,0,l_1,l_2)$ on the left and the amplitudes $ (l_3,l_4,0,0)$; in either side, only exponentially decaying modes are allowed. This results in the equation to be solved for the energy of the bound states as
		\begin{equation}
			\mathcal{T}_{34} \mathcal{T}_{43}=\mathcal{T}_{33} \mathcal{T}_{44}.
		\end{equation}
		Using this procedure, we compute the bound states that are supported in a system of two moir\'e periods, which is presented as a function of the inter-chain coupling $V$ in Fig.~\ref{fig:boundstates}. One observes that flat band transitions happen very close to the degenerate point of the two bound states of the system of two moir\'e periods.
		\begin{figure}
			\begin{center}
				\includegraphics[width=0.5\textwidth]{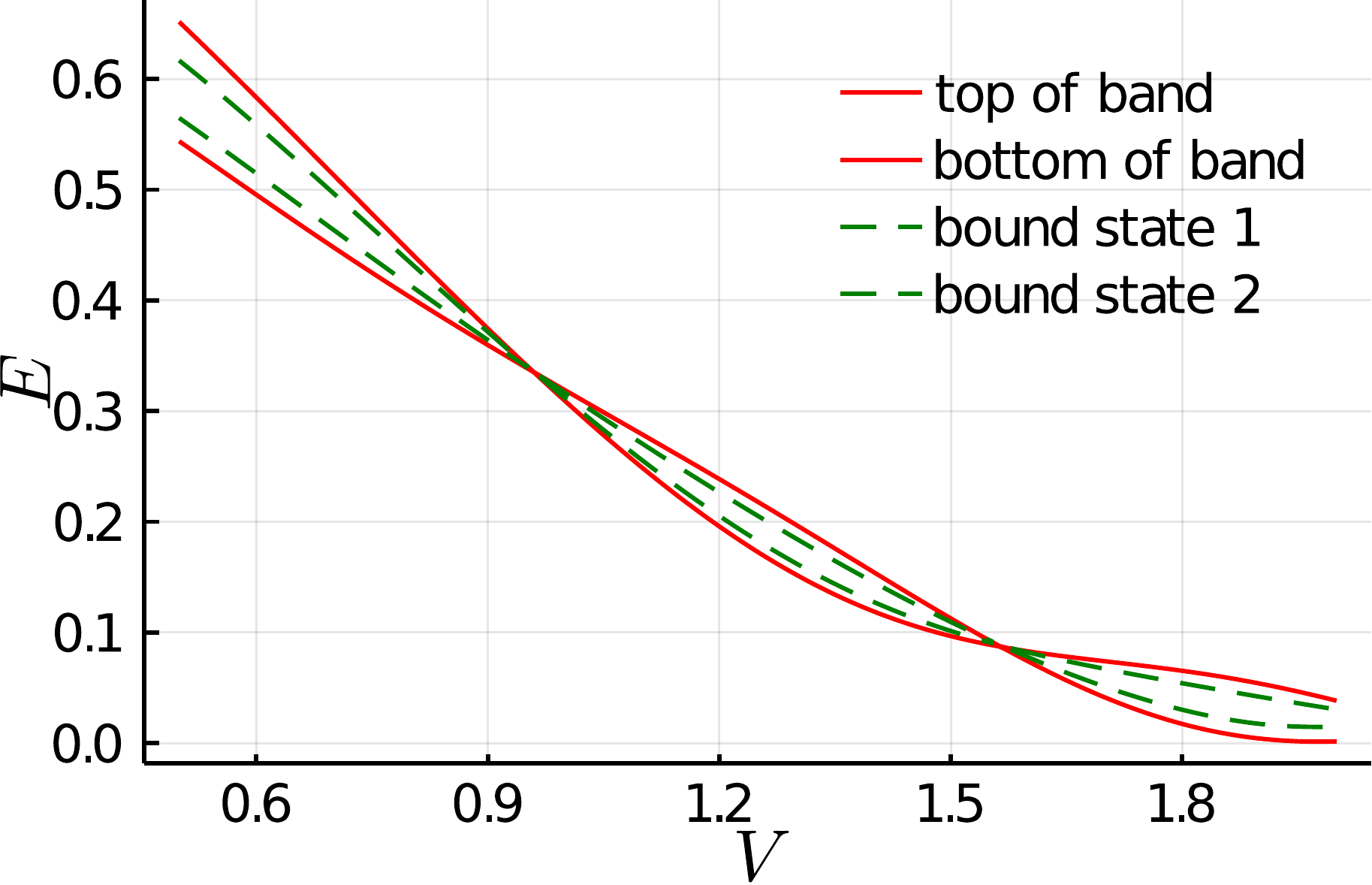}
			\end{center}
			\caption{Comparison of bound states of two moir\'e periods with the top and the bottom of the lowest positive band of the periodic system. Flat bands happen close to the degenerate point of the two bound states of the system of two moir\'e periods. Here $U=1$, $\Delta=-0.3$, $q_0=0.6$.}
			\label{fig:boundstates}
		\end{figure}
		\subsubsection{Derivation of the band structure of the infinite system}

		As an interesting side remark, we mention that the band structure of the system can also be computed from the generalised transfer matrix $G(x_2,x_1)$. To this end, we choose $x_2-x_1$ to be an apparent period of the potential (twice as much of the moir\'e period), $x_2-x_1 = 2 \Lambda$. Then from the fact that $\Psi(x_2) = G(x_2,x_1) \Psi(x_1)$ and the Bloch theorem $\Psi(x_2)= e^{iq 2 \Lambda} \Psi(x_1)$ we obtain $\det [G (x_2,x_1) - e^{i q 2 \Lambda}]=0$. This allows one to compute the Bloch wavevector corresponding to the energy under consideration $E$. By selecting the real wave vector $q$, the band structure of the system can then be derived. 
		
		\section{Parameter retrieval for the effective Hamiltonians}\label{sec:parameter_retrieval}
		
		The effective Hamiltonian of the moir\'e structure is determined by the energies $\omega_0^{(1,2)}$, $U^{(1,2)}$, $V$ and the group velocity $v$. These values are retrieved from the  dispersion characteristics of the single layer structure (for $\omega_0$,$U$ and $v$), and of the bilayer structure (for $V$) which are obtained by RCWA simulations. In the following, we will discuss in details these parameter retrieval methods. 
		
		\subsection{Parameter retrieval of single grating structure}\label{sec:parameter_retrieval_single}
		
		The dispersion characteristic of a single grating structure is easily calculated from Eq.~ \eqref{eq:Hsingq} in the main text. It consists of two bands of opposite curvature $\pm \frac{v^2}{2U}$, with corresponding band edge energies given by $\omega_0 \pm U$. As a consequence, $\omega_0$ and $U$ are directly extracted from the energy of resonances at $q=0$ of the RCWA simulations. Then knowing $U$, the group velocity $v$ is extracted from the curvature of these resonance. As shown in Fig.~\ref{fig:parameter_single}b, the band structure which is calculated by the effective Hamiltonian using the retrieved parameters reproduce perfectly the simulated one.
		
		\begin{figure}[hbt!]
			\begin{center}
				\includegraphics[width=1\textwidth]{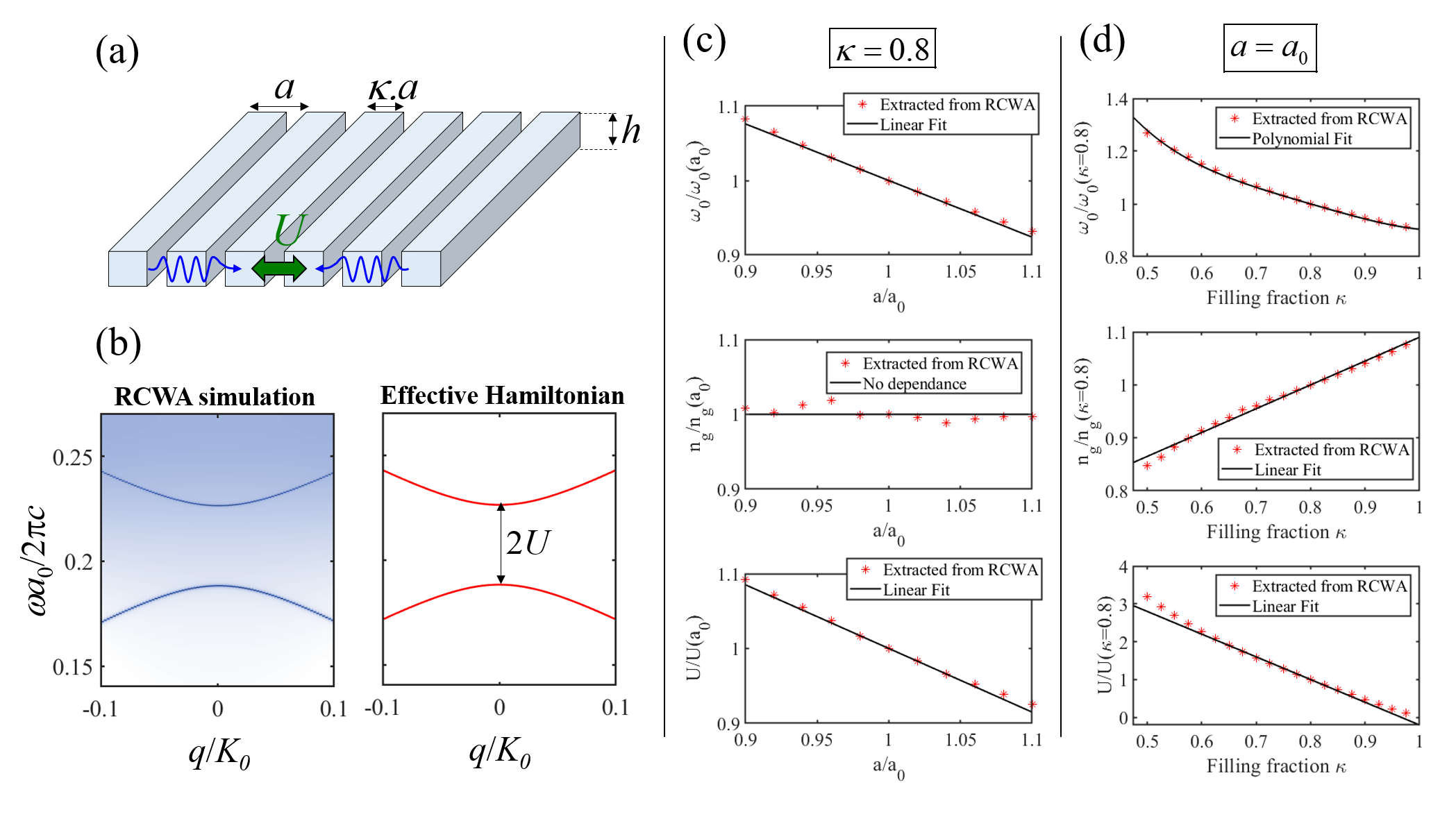}
			\end{center}
			\caption{(a) Sketch of a single grating structure. (b) Band structure of a single grating structure obtained by RCWA simulation (left) and by the effective Hamiltonian using retrieved parameter $U,\omega_0,v$. Here the simulated structure corresponds to $a=a_0, \kappa=0.8$ and $h=0.6a_0$. The retrieved parameters are $\omega_0=\Omega_0$, $U=U_0$  and $n_g=c/v=3$. With $\Omega_0a_0/2\pi c=0.2073$ and $U_0a_0/2\pi c=0.0191$.(c) Dependence of the retrieved parameters when the period $a$ is slightly difference than $a_0$. It shows that the group velocity $v$ is almost unchanged, while $\omega_0$ and $U$ are slightly modified. The modifications of $\omega_0$ and $U$ can be fitted by $\omega_0(a)=\Omega_0\left[1-0.76\left(a/a_0-1\right)\right]$ and $U(a)=U_0\left[1-0.85\left(a/a_0-1\right)\right]$. (d) Dependence of the retrieved parameters when the filling fraction $\kappa$ is scanned from 0.5 to 1. It shows that while the offset energy $\omega_0$ and the group velocity are slightly modified, the intra-layer coupling strength $U$ is greatly modified from 3$U_0$ to 0.} \label{fig:parameter_single}.
		\end{figure}
		
		With the retrieval method presented above, we can explore the dependence of $U$, $\omega_0$ and $v$ on geometrical parameters of the system. In particular, two dependencies are studied in details: 
		\begin{itemize}
			\item {\it\textbf{Dependence on the period $a$ when $a$ is slightly different than $a_0$}}:  this dependence is responsible to the slight difference between $U^{(1)},\omega_0^{(1)}$ and $U^{(2)},\omega_0^{(2)}$ corresponding to upper and lower gratings of period $a_1$ and $a_2$. The results of this study are shown in Fig.~\ref{fig:parameter_single}c. We notice that the linear dependence $\omega_0(a)$ leading to a simple proportional relation between  $\Delta=\frac{\omega_0^{(1)}-\omega_0^{(2)}}{2}$, $\Delta_U=\frac{U^{(1)}-U^{(2)}}{2}$ and $\frac{1}{N} \approx \frac{a_2-a_1}{a_0}$. As a consequence, the three parameters $\Delta$,$\Delta_U$ and $q_0$ of the Hamiltonian \eqref{eq:simplified_MoireH} are connected and can be reduced to a single one, for example $q_0$.  
			\item {\it\textbf{Dependence on the filling fraction $\kappa$}}: the strong and almost linear dependence of $U(\kappa)$ is shown in Fig.~\ref{fig:parameter_single}d. It suggests that the filling fraction is the parameter for tuning the intralayer coupling strength. 
		\end{itemize}
		
		\subsection{Parameter retrieval of bilayer structure}\label{sec:parameter_retrieval_bi}
		
		\begin{figure}[hbt!]
			\begin{center}
				\includegraphics[width=0.85\textwidth]{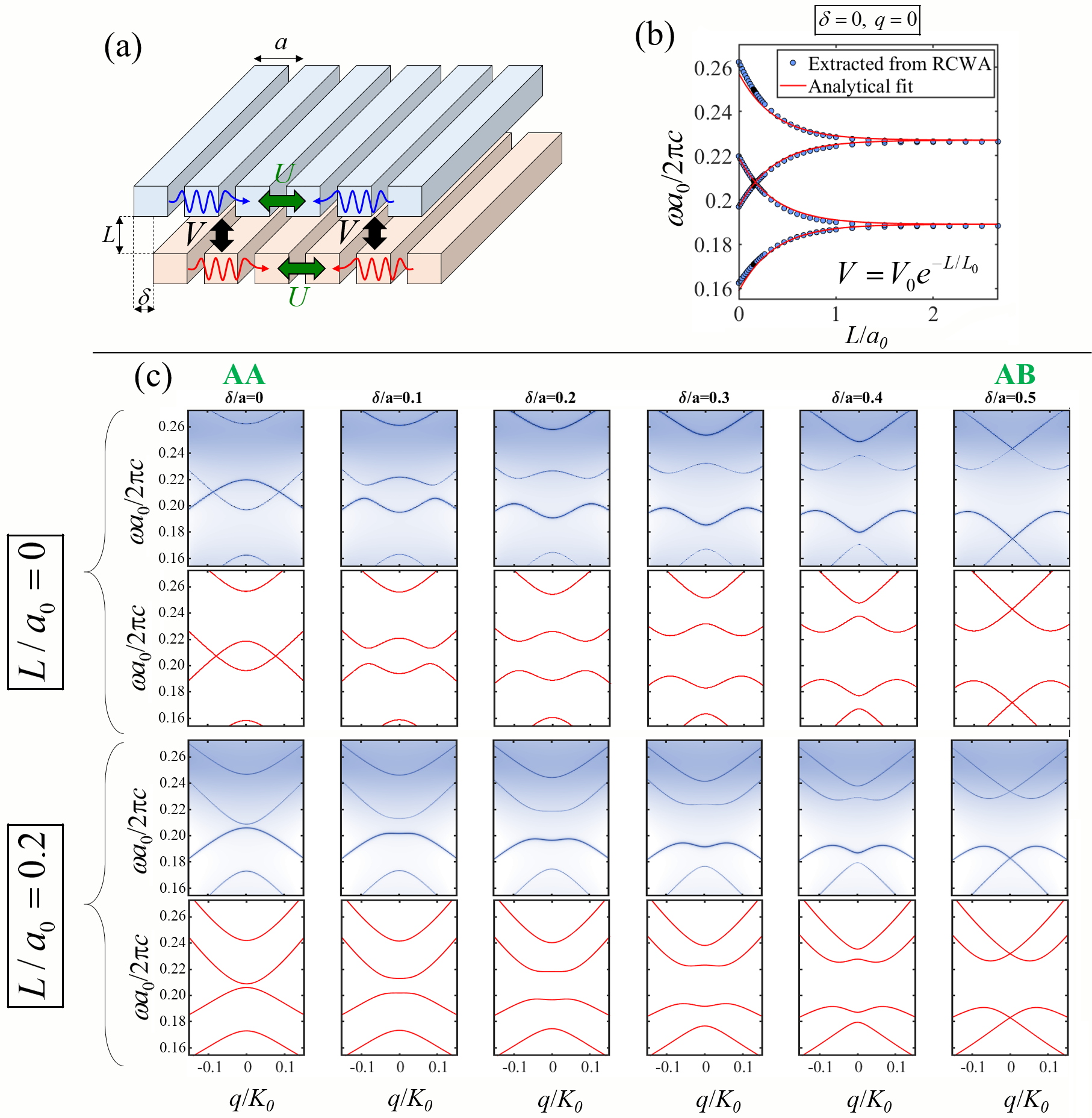}
			\end{center}
			\caption{(a) Sketch of a bilayer grating structure. (b) Band-edge energies of the band structure of a bilayer grating as a function of the distance $L$ between the two layers. The two grating are identical and aligned, with $a=a_0$, $\kappa=0.8$ and $h=0.6 a_0$. The blues circles correspond to extracted data from RCWA simulation. The solid red lines are fittings, given by $\omega_0 \pm U - V$ and $\omega_0 \pm U + V$. Here $\omega_0=\Omega_0$ and $U=U_0$, obtained from parameter retrieval of the single grating. And $V(L)=V_0e^{-L/L_0}$ with $V_0a_0/2\pi c=0.032$ and $L_0/a_0=0.34$. (c) Band structure of bilayer grating structures of different relative displacement $\delta/a$, obtained by RCWA simulation  and by the effective Hamiltonian using retrieved parameter $U,\omega_0,v$ and $V$.}
			\label{fig:parameter_fishbone}
		\end{figure}
		
		The dispersion characteristic of bilayer structure can be analytically calculated from Eq.~\eqref{eq:bilayer} from the main text. The detailed of these eigenmodes has been reported in \cite{NHS2018}. Here we only discuss how to retrieve the inter-layer coupling strength from these band structure and the validation of the method. \\
		
		Since $\omega_0$ and $U$ are already retrieved from the simulation of single grating, only $V$ left to be retrieved. One may show that, for $AA$ stacking (i.e. $\delta/a=0$), the band structure consist of four bands with bandedge energies given by $\omega_0\pm U + V$ and $\omega_0 \pm - V$. As a consequence, $V$ is directly extracted from the energy of resonance at $q=0$ of the RCWA simulations for anyone from the four bands. Using this method, we can easily obtain the dependence of $V$ as the function of the distance $L$ separating the two grating. The results shown in Fig.~\ref{fig:parameter_fishbone}b evidences the dependence law $V=V_0e^{-L/L_0}$ used in the main text.\\
		
		Finally, we confirm the validity of the retrieved parameters by using them to calculate the band structure of the bilayer for different relative shift $\delta/a$, and for diffrent value pof $L$. The results presented in Fig.~\ref{fig:parameter_fishbone}c show perfect aggreement between the calculated dispersion and the ones obtained by RCWA simulations, thus validate the retrieved parameters.
		
		\section{Band edges of moir\'e bands}\label{sec:bandeges}
		\begin{figure}[hbt!]
			\begin{center}
				\includegraphics[width=0.7\textwidth]{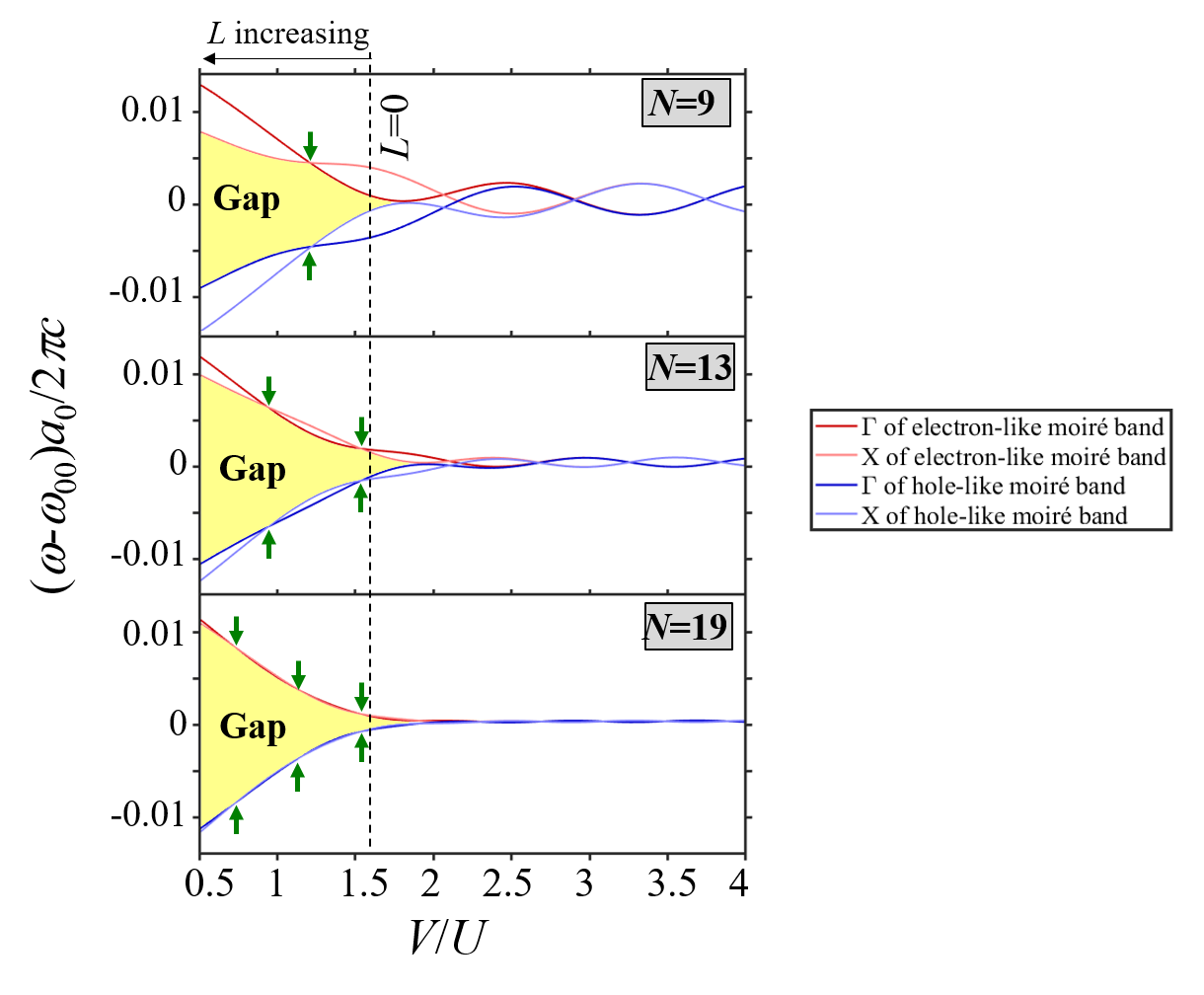}
			\end{center}
			\caption{Results from the effective Hamiltonian of the band-edge energies (at $\Gamma$ and $X$ points of electron-like and hole-like moir\'e band) when scanning $V/U$ for different moir\'e structures. The vertical black dashed line indicates the value of $V/U$ corresponding to $L=0$. The green arrows indicate flat band configurations.}
			\label{fig:bandegdes}
		\end{figure}
		
		To investigate the interplay between intra and inter-layer coupling in the formation of moir\'e bands, the band-edge energies (at $\Gamma$ and $X$ points) of electron-like and hole-like moir\'e bands are extracted from effective Hamiltonian calculations when scanning the ratio $V/U$ for different moir\'e configurations with fixed value of $U=U_0$. The results depicted in Fig~\ref{fig:bandegdes} evidence two important features:
		\begin{itemize}
			\item The magic configuration takes place at the crossings of band edge energies from the same miniband when tuning $V/U$ (indicated by green arrows in  Figs~\ref{fig:bandegdes}).
			\item The two moir\'e bands get closer when increasing $V/U$, as previously discussed when scanning $L$ in the maintext. Interestingly, the gap between them is closed, and they merge together when $V/U \gtrsim 2$ for all value of $N$. Indeed, the bandgap when the two gratings are slightly different (i.e. $N\gg 1$) and uncoupled (i.e. $V\ll U$) is given by the gap of a single grating, thus amounts to 2$U$. When the interlayer layer coupling $V$ is implemented, the two moir\'es bands emerge and are separated to the corresponding continuum by a quantity  $\sim V$. Thus they would merge at the zero energy when $V\sim U$  . This feature is not revealed from the numerical simulation since the maximum value of $V/U$ from our design is 1.76 (i.e. $L=0$ and $\kappa=0.8$).
		\end{itemize}
	\end{widetext}
\end{document}